\documentclass[aps,prb,twocolumn,floats,epsfig,showpacs]{revtex4-1}

\usepackage[apple mac]{inputenc}
\usepackage{fourier}
\usepackage{textcomp}

\usepackage{bbm}
\usepackage{amssymb}
\usepackage{ifpdf}
\usepackage{color}
\usepackage{graphicx}
\usepackage{amsmath}
\usepackage{amsfonts}
\usepackage{dcolumn}
\usepackage{bm}
\usepackage{xcolor}

\begin{document}

\title{Efficient algorithm to compute the Berry conductivity}

\author{A. Dauphin$^{1,2}$, M. M\"uller$^{1}$, and M. A. Martin-Delgado$^{1}$}
\affiliation{$^1$Departamento de Fisica Teorica I, Universidad Complutense de Madrid, 28040 Madrid, Spain \\
$^2$Center for Nonlinear Phenomena and Complex Systems, Universit\'{e} Libre de Bruxelles (U.L.B.), Code Postal 231, Campus Plaine, B-1050 Brussels, Belgium}

\begin{abstract}
We propose and construct a numerical algorithm to calculate the Berry conductivity in topological band insulators. The method is applicable to cold atom systems as well as solid state setups, both for the insulating case where the Fermi energy lies in the gap between two bulk bands as well as in the metallic regime and interpolates smoothly between both regimes. The algorithm is gauge-invariant by construction, efficient and yields the Berry conductivity with known and controllable statistical error bars. We apply the algorithm to several paradigmatic models in the field of topological insulators, including Haldane's model on the honeycomb lattice, the multi-band Hofstadter model and the BHZ model, which describes the 2D spin Hall effect observed in CdTe/HgTe/CdTe quantum well heterostructures.
\end{abstract}


\maketitle



\section{\label{sec:introduction}Introduction}

Topological insulators (TI) are a topological state of quantum matter which constitutes a new paradigm in condensed matter physics \cite{hasan-rmp-82-3045, qi-physics-today-2010, moore-nphys-5-378, qi-rmp-83-1057}. These recently discovered new materials exhibit unique fascinating properties such as current-carrying surface and edge states that are strongly protected against perturbations in either the bulk or the surface of the material \cite{fu-prl-98-106803, xia-natphys-5-398, chen-science-325-178, zhang-natphys-5-438, wray-natphys-7-32, hsieh-prl-103-146401} and non-standard exchange statistics of quasi-particle excitations, which offer potential applications in the context of quantum computation \cite{fu-prl-100-096407, qi-science-323-1184, nayak-rmp-80-1083}.

The question what happens in topological insulators when the Fermi energy does no longer lie inside the gap between two energy bands, is by no means rhetoric but of high practical importance: in fact, this situation naturally occurs in the experimental process of production of candidate samples of topological insulators such as Bi$_2$Se$_3$ and Bi$_2$Te$_3$ compounds. These are used for instance in cooling devices due to their favorable thermoelectric properties. The chemical composition can be well-controlled and adjusted to the composition of the desired topological insulator. However, it is much more demanding to control the level of the Fermi energy, which for many samples lies within the bulk energy bands instead of the insulating energy gap, thereby invalidating them as true TIs. This difficulty has motivated the development of sophisticated molecular beam epitaxy (MBE) techniques to precisely control the growth of ultra-thin Bi$_2$Se$_3$ and Bi$_2$Te$_3$ films \cite{lee-appl_phys_lett-101-013118, cao-appl_phys_lett-101-162104}. Likewise, in two-dimensional TIs it is possible to adjust the Fermi energy to lie either in the band gap or the bulk bands. Experimentally, in CdTe/HgTe/CdTe quantum wells, formed by a thin layer of HgTe embedded between two CdTe layers, this can be achieved by an elaborated MBE technique that allows one to control the thickness of the intermediate HgTe layer and thereby tune the position of the Fermi energy with respect to the bands \cite{bernevig-science-314-1757, koenig-science-318-766}. For an appropriate thickness, the Fermi energy lies in the gap between the bulk bands and the heterostructure shows the desired characteristic topological insulating behavior with a quantized spin conductivity of $2 e^2/h$. 

Complementary to solid-state realizations, cold atoms in optical lattices have been proposed as a realistic platform to experimentally explore the new physics of TIs under controllable conditions \cite{Goldman:2013p1721,Mazza-njp-14-015007, hauke-prl-109-145301, lang-prl-108-220401, kraus-prl-109-106402, mei-pra-85-013638, zhu-prl-97-240401, umucalilar-prl-100-070402, stanescu-pra-79-053639, goldman-prl-105-255302, stanescu-pra-82-013608, bermudez-prl-105-190404, beri-prl-107-145301, goldman-njp-15-013025, goldman-jphysb-46-134010, goldman-pnas-110-6736, alba-prl-107-235301}. In particular, in these systems the Fermi energy can be controlled directly by the filling of atoms in the lattice and there are several proposals to measure the transverse conductivity for both the insulating and the metallic case \cite{alba-prl-107-235301,goldman-njp-15-013025,hauke-prl-109-145301}. In contrast, in condensed matter systems such as the above-mentioned chemical compounds the pinning of the Fermi level to a value inside the bulk bands typically arises due to external causes like crystal defects and other sources which are not straightforward to control. As a consequence, in transport properties and measurements bulk carriers often dominate over the contribution stemming from surface or edge states. 

\begin{figure}[t]
\begin{center}
	\includegraphics[scale=1]{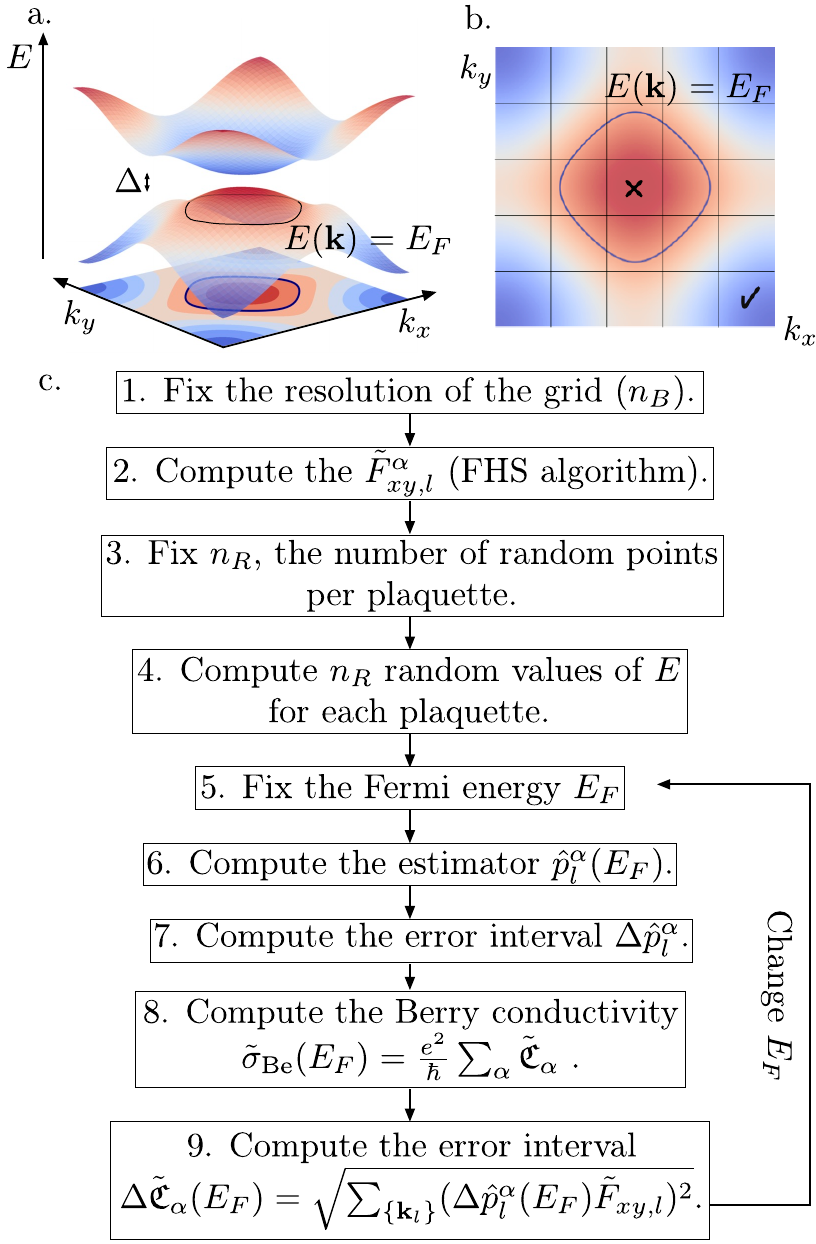}
	\caption{\textbf{a.} Generic energy spectrum of a system with an energy gap $\Delta$. In the displayed situation the Fermi energy falls into the first energy band and defines the Fermi surface as the equipotential energy line at $E_\alpha(\mathbf{k}) = E_F$ (solid line). The projection of the energy dispersion of the first band is shown as a color-coded plot in the horizontal $k_x - k_y$ -- plane. \textbf{b.} For the numerical calculation of the Berry conductivity, the Brillouin zone is discretized by a finite grid. Momentum space plaquettes with energies $E_\alpha (\mathbf{k})$ entirely below (above) the Fermi energy contribute entirely (not at all) to the Berry conductivity, whereas plaquettes which cut the Fermi surface contribute partially. \textbf{c.} Schematic summary of the numerical algorithm to calculate the Berry conductivity: After fixing the discretization grid of momentum space and calculating the Berry curvature contributions by means of the FHS algorithm for each plaquette of the Brillouin zone, a classical Monte Carlo sampling method is used to determine the weights with which the individual plaquettes contribute to the conductivity. Statistical uncertainties in the sampling process result in controlled and statistical errors in the Berry conductivity.}
\label{fig:fig1}
\end{center}
\end{figure} 

Finally, this question plays as well a fundamental role in the physics of the anomalous quantum Hall effect (AHE) \cite{hall-Philos_mag-12-157, nagaosa-rmp-82-1539}, which precedes the upsurge of topological insulators as a prominent field in condensed matter. In the standard quantum Hall effect (QHE), which can be observed in non-magnetic materials, there is a linear dependence of the Hall resitivity $\rho_{xy}$ on an externally applied perpendicular magnetic field. In contrast, in the AHE an anomalous deviation from the linear law is observed in ferromagnetic materials. A complete theory for the AHE has remained elusive for more than a century, largely due to the complications arising from the fact that there are three main mechanisms that influence the electronic motion and can give rise to an AHE: the intrinsic mechanism, the skew-scattering mechanism and the side-jump mechanism\cite{nagaosa-rmp-82-1539}. Here, we shall be interested in the so called intrinsic mechanism for the AHE, which is the contribution that can be expressed in terms of the Berry-phase curvature and thereby represents an intrinsic quantum mechanical property of a perfect crystal. This intrinsic contribution, which is dominant in metallic ferromagnets with moderate conductivity, depends only on band structure properties and is largely independent of scattering that affects other AHE mechanisms. 

Understanding of this intrinsic and anomalous contribution has become possible with the seminal work by Haldane \cite{haldane-prl-93-206602} who uncovered by a fully quantum-mechanical treatment, unlike precedent work based on semiclassical methods \cite{karplus-pr-95-1154}, the topological origin of this contribution and its relation to the physics taking place at the Fermi surface. Haldane showed that the intrinsic contribution to the AHE conductivity stems from a combination of an integer-valued part stemming from the contribution of filled bands and a part originating from the Fermi surface, i.e.~from the cuts of a partially filled band at the Fermi energy $E_F$ (non-integer valued contribution). 

It is crucial to realize that in order to directly apply Haldane's equations \cite{haldane-prl-93-206602} to a given problem, one needs to know precisely the form of the Fermi surface. In practice, except in very simple model cases, this is not possible since the band structure of real materials is obtained from detailed numerical calculations and one is typically given a numerical data set about the bands instead of an explicit formula. Thus, in practice it is highly desirable to have at one's disposal a numerical method, which is (i) gauge-invariant, (ii) efficient and (iii) outputs numerical results with controllable and known error intervals. In this work, we develop such a general and efficient numerical algorithm to compute the Berry conductivity when the Fermi energy does not lie within the band gap. In the following, we shall refer to Berry conductivity as the non-quantized conductivity associated to the Chern number according to the according to the Thouless, Kohmoto, Nightingale, Den Nijs (TKNN) formula \cite{thouless-prl-49-405} when the position of the Fermi level lies in the conduction band, so that we recover the TKNN quantized conductivity for the standard insulating case if the Fermi energy lies in the energy gap between two bands.

\vspace{1mm}
Our main results are:
\vspace{1mm}

\noindent i/ We present a new method to compute the Berry conductivity when the Fermi energy level is located outside the band gap. We outline the algorithm (schematically summarized in Fig.~\ref{fig:fig1}), discuss its ingredients and show that it is gauge invariant and efficient (Sec.~\ref{sec:method}). 
\vspace{1mm}

\noindent ii/ We emphasize that a central feature of the presented method is that it is endowed with known and controllable error bars for the non-integer value of the conductivity. This is essential. When the Berry conductivity is not integer-valued, errors due to approximations need to be under control in order to distinguish two different values of the conductivity observable, so that one safely distinguish a topological phase from a trivial phase.
\vspace{1mm}

\noindent iii/ To test and benchmark the performance of the algorithm we first apply it to the paradigmatic Haldane model \cite{haldane-prl-61-2015}, which has a simple enough structure so that the analytic form of the two-band energy spectrum is known (Sec.~\ref{sec:models:haldane}). Subsequently, we apply the method to the more complex case of the Hofstadter model \cite{hofstadter-prb-14-2239}, which belongs to the class of multi-band topological insulators, where the band structure information is obtained numerically (Sec.~\ref{sec:models:hofstadter}). These models are both of importance and have attracted interest in the field of quantum simulation of topological insulators with cold atoms in optical lattices. Here, our method provides the theoretical tools that allow one to map out the phase diagrams in future experiments. Finally, we also apply our method to the BHZ model \cite{bernevig-science-314-1757} which is a realistic model which captures the physics of 2D spin Hall effect present in systems such as the above-mentioned CdTe/HgTe/CdTe quantum well compounds (Sec.~\ref{sec:models:bhz}). We conclude with a short summary and a discussion of possible future extensions of the presented method (Sec.~\ref{sec:conclusions_and_outlook}).

\section{\label{sec:method}Conceptual Outline of the Algorithm}

\subsection{\label{sec:generalized_Berry}Generalized Berry conductivity}
Before presenting our numerical algorithm to calculate the Berry conductivity, in this section we briefly review the expressions for the intrinsic Hall conductivity both for the insulating case where the value of the Fermi energy lies in the gap between two bands, as well as the generalized result for the situation in which the Fermi energy lies in a partially filled band \cite{haldane-prl-93-206602}. 

In the insulating case, the Hall conductivity is quantized and proportional to the sum of the Chern numbers of the occupied energy bands, 
\begin{equation}
\label{eq:quantisedcond}
\sigma_H=\frac{e^2}{h}\sum_{E_\alpha<E_F} C_\alpha. 
\end{equation}
The Chern numbers $C_\alpha$ are integer-valued topological invariants, defined in terms of the integral of the Berry curvature $F^\alpha_{xy}(\mathbf{k})$ over the whole Brillouin Zone (B.Z.) \cite{thouless-prl-49-405, kohmoto-ann-phys-160-343}:
\begin{equation}
\label{eq:chernnumber}
\begin{split}
&C_\alpha=\frac{1}{2\pi i}\int_{B.Z.}F^{\alpha}_{xy}(\mathbf{k}) d^2k\\
&F^\alpha_{xy}(\mathbf{k})=\partial_{k_x}A^\alpha_y(\mathbf{k})-\partial_{k_y} A^\alpha_x(\mathbf{k})\text{.}
\end{split}
\end{equation} 
The latter is expressed by the exterior derivative of the Berry connection
\begin{equation}
\label{eq:berryconnection}
A^\alpha_\mu (\mathbf{k})=\langle u_\alpha(\mathbf{k})|\partial_\mu | u_\alpha(\mathbf{k})\rangle\text{,}
\end{equation}
where $u_\alpha(\mathbf{k})$ is the eigenvector corresponding to the energy band $E_\alpha(\mathbf{k})$.

In the case that the Fermi energy does not lie in an energy gap between bands, as schematically shown in Fig.~\ref{fig:fig1}a, the intrinsic Hall conductivity generalizes to \cite{nagaosa-rmp-82-1539,xiao-rmp-82-1959}
\begin{equation}
\label{eq:Berry_Conductivity}
\sigma_H(E_F)=\frac{e^2}{h}\sum_{\alpha}  \mathfrak{C}_\alpha \text{,}
\end{equation}
with
\begin{equation}
\label{eq:partialchern}
\mathfrak{C}_\alpha(E_F)=\frac{1}{2\pi i}\int_{B.Z.} d^2k \, F^{\alpha}_{xy}(\mathbf{k}) \Theta(E_F-E_\alpha(\mathbf{k})) \text{,}
\end{equation}
where $\Theta(E)$ denotes the Heaviside function and $\alpha$ denotes the band index. Thus, the conductivity is the sum of the integer-valued Chern numbers corresponding to fully-occupied energy bands below the Fermi energy and a non-quantized contribution which depends on the Fermi surface, i.e.~it stems from the integral over energy band(s), which are partially filled at a given Fermi energy $E_F$. 

For systems with a particularly simple band structure, as e.g.~in two-band systems, the expressions for the eigenvalues and eigenvectors of the bands are given in explicit form, and hence the Chern values $\mathfrak{C}_\alpha$ can be calculated analytically. In general, however, the system Hamiltonian cannot be diagonalized analytically and an efficient numerical method to compute the Chern values is needed.

\subsection{\label{sec:algorithm}Construction and properties of the algorithm}

The algorithm we propose to numerically compute the Chern values of Eq.~(\ref{eq:partialchern}) and thereby the Berry conductivity of Eq.~(\ref{eq:Berry_Conductivity}) is based on a series of controlled approximations: First, we discretize the two-dimensional Brillouin zone by a finite $n_B \times n_B$ grid of small plaquettes at discrete momenta $\mathbf{k}_l$ (see Fig.~\ref{fig:fig1}b and Appendix \ref{app:fukui} for details), so that the integral over the (partially filled) band becomes 
\begin{equation}
\label{eq:partialchern_discr}
\mathfrak{C}_\alpha (E_F) \longrightarrow \frac{1}{2\pi i}\sum_{\{\mathbf{k}_l \}} F^{\alpha}_{xy, l} \, p^{\alpha}_l(E_F)
\end{equation}
with the Berry curvature contribution 
\begin{equation}
F^{\alpha}_{xy, l} = \int_\square d^2k \, F^{\alpha}_{xy}(\mathbf{k})  
\end{equation}
from a small two-dimensional plaquette of size $\Delta_{k_x} \Delta_{k_y}$, and the weighting factors
\begin{equation}
p^\alpha_l (E_F) = \frac{1}{\Delta_{k_x} \Delta_{k_y}} \int_\square d^2k \, \Theta(E_F-E_\alpha(\mathbf{k})).
\end{equation}
The weights $p^\alpha_l(E_F)$ correspond to the partial area of the plaquette, which is covered by the Fermi sea, thus $p^\alpha_l(E_F) = 0$ ($p^\alpha_l(E_F) = 1$) for squares with energies completely above (below) the Fermi energy $E_F$, and $0 < p^\alpha_l(E_F) < 1$ for momentum space plaquettes which are cut by the Fermi surface (see Fig.~\ref{fig:fig1}b). The choice the value $n_B$, i.e.~the resolution of the momentum space grid, is important: it can be motivated either by given physical conditions, such as a finite experimental energy resolution or e.g.~the finite size of real-space optical lattices, which in turn induces a smallest characteristic scale in momentum space; or it can be chosen according to given numerical resources. In Appendix \ref{app:conv}, we derive a convergence criterion in terms of the grid discretization and provide an error bound due to the grid discretization.

The key of the numerical algorithm is now to evaluate reliably and under controlled approximations the discretized sum of Eq.~(\ref{eq:partialchern_discr}), whose value converges to Eq.~(\ref{eq:Berry_Conductivity}) for increasingly finer grids. 

\textit{(i) Gauge-invariant calculation of the Berry curvature:} To numerically calculate the Berry curvature contributions $F^{\alpha}_{xy, l}$ we employ a numerical algorithm proposed by Fukui, Hatsugai and Suzuki \cite{bib:fukui} (FHS algorithm). It is highly efficient and the discrete sum $1 / (2\pi i) \sum_{\{\mathbf{k}_l \}} \tilde{F}^{\alpha}_{xy, l}$ converges rapidly to the correct integer-valued Chern numbers $C_\alpha$, even for a very coarse-grained discretization of the Brillouin zone. This behavior is rooted in the fact that the algorithm is based on a lattice gauge formulation\cite{wilson-prd-10-2445, phillips-annphys-prd-161-399} instead of a finite difference discretization of the Berry curvature. In Appendix \ref{app:fukui} we provide a brief summary of the FHS algorithm and the explicit expressions for the lattice strength $\tilde{F}^{\alpha}_{xy, l}$ calculated with the FHS method. 

\textit{(ii) Efficient estimation of the weights $p^\alpha_l(E_F)$:} To decide whether a given plaquette in momentum space contributes entirely, partially or not at all, we use a simple and rapid classical Monte-Carlo technique: for each plaquette of the grid localized around the discrete momentum $\mathbf{k}_l$, we generate $n_R$ uniformly distributed random points $\mathbf{k}_R$ within the plaquette and compute $E_\alpha(\mathbf{k}_R)$ which lies above or below the Fermi Energy. Based on the latter we define the estimator 
\begin{equation}
\label{eq:probareaest}
\hat{p}^\alpha_l (E_F) = \frac{1}{n_R} \sum_{\{\mathbf{k}_R\}} \Theta(E_F-E_\alpha(\mathbf{k}_R))
\end{equation}
for the weighting factors $p^\alpha_l(E_F)$.  

\textit{(iii) Statistical confidence interval and controlled numerical error of the Berry conductivity:} Note that the randomness of this estimation procedure introduces a statistical uncertainty. Note that the value of the estimators $\hat{p}^\alpha_l(E_F)$ is bounded between zero and one. However, it is clear that the statistical error will be largest for partially contributing plaquettes with $\hat{p}^\alpha_l(E_F) \sim 1/2$, whereas the uncertainty in $\hat{p}^\alpha_l(E_F)$ for plaquettes with energies completely above or completely below the Fermi energy is expected to be much smaller. In order to have a known and minimal statistical error in $\hat{p}^\alpha_l(E_F)$, and thus in the Berry conductivity, it is highly desirable that the numerical algorithm takes this effect into account and provides statistical errors which depend on the actual value of the Fermi energy. The quantity $\hat{p}^\alpha_l(E_F)$ is the estimator of the fixed though unknown parameter $p$ of a binomial distribution $\mathcal{B}(n_R, p)$, corresponding to the process of tossing $n_R$ times a biased coin. As is discussed in detail in Appendix \ref{app:detailsalgo}, using the normal approximation and for a fixed number of runs $n_R$ and a desired value $\epsilon < 1$ this allows one to derive a confidence interval $[p^\alpha_{l, \text{min}},\, p^\alpha_{l, \text{max}}]$ for $\hat{p}^\alpha_l(E_F)$, called the \textit{Wilson interval} \cite{brown-stat-science-16-101} with modified boundary conditions. This means that with a probability $1 - \epsilon$ the "true" value $p^\alpha_l$ lies in this interval. The key point is that the width of this interval depends on the actual value of the estimator $\hat{p}^\alpha_l$ and is typically significantly smaller than the trivial upper bound of one. After symmetrizing the interval by taking the maximum $\Delta \hat{p}^\alpha_l(E_F) = \text{max}(\hat{p}^\alpha_l-p_\text{min},p_\text{max}-\hat{p}^\alpha_l)$, each momentum space plaquette of the grid is associated to a probability value $\hat{p}_l (E_F) \pm \Delta \hat{p}_l(E_F)$ with a confidence of at least $1-\epsilon$. Finally, we remark that whereas the discussed statistical method is conceptually simple, intuitive and provides controllable error bars, it could be refined and combined with more sophisticated techniques to evaluate the weighting factors (\ref{eq:probareaest}) or, equivalently, to determine the equal-energy contours of the bands for a given Fermi energy. In addition, an adaptative version of the statistical algorithm, in which only weighting factors of plaquettes with large Berry curvature contributions are evaluated with high statistical accuracy, could be put forward. 

\begin{figure}[t]
\begin{center}
	\includegraphics[scale=1]{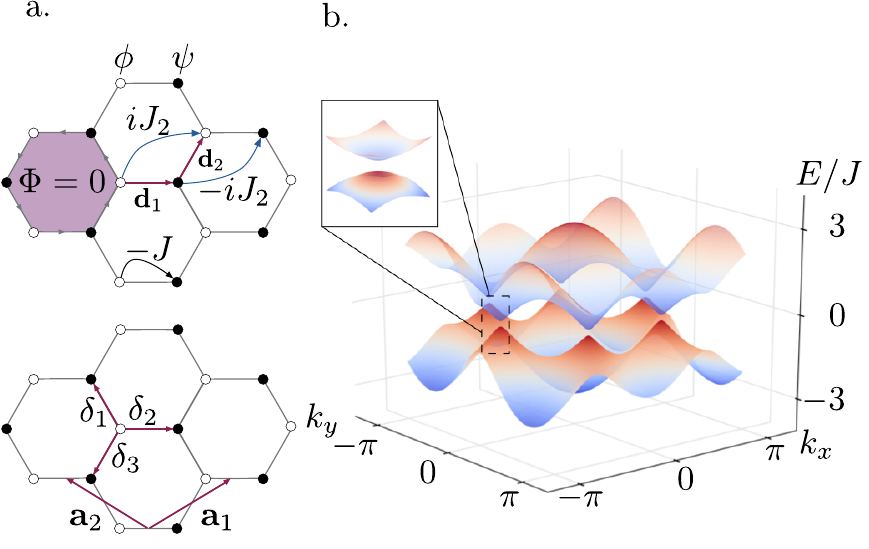}
	\caption{Haldane model of spinless fermions on the honeycomb lattice: a. Dynamics is governed by a real-valued nearest-neighbor hopping and an imaginary-valued next-to-nearest neighbor hopping amplitude, in combination with a staggering potential which induces a chemical potential difference between $\phi$- and $\psi$-lattice sites. The net magnetic flux $\Phi$ through a unit cell is null. Fig.~b. shows the energy spectrum for $J_2=0.1J$ and $\beta=0$: the imaginary N.N.N. hopping opens a topologically non-trivial gap at the two inequivalent Dirac cones.}
\label{fig:fig2}
\end{center}
\end{figure} 

As mentioned above, for even moderately fine grids the FHS algorithm provides essentially exact values for the Berry curvature contributions (see \cite{bib:fukui} and Appendix \ref{app:fukui}). Thus, the statistical uncertainty of $\hat{p}^\alpha_l$ directly translates into an uncertainty in the Berry conductivity contributions, 
\begin{equation}
p^\alpha_l(E_F) F^{\alpha}_{xy, l} \longrightarrow \hat{p}^\alpha_l (E_F) \tilde{F}^\alpha_{xy,l}\pm \Delta \hat{p}^\alpha_{l} (E_F) \tilde{F}^\alpha_{xy,l}\text{.}
\end{equation}
Finally, the estimated Berry conductivity is given by 
\begin{equation}
\label{eq:Berry_Conductivity_estim}
\tilde{\sigma}_\text{Be}(E_F)=\frac{e^2}{h}\sum_{\alpha}  \tilde{\mathfrak{C}}_\alpha(E_F),
\end{equation}
where
\begin{equation}
\label{eq:partialchern}
\tilde{\mathfrak{C}}_\alpha(E_F) = \frac{1}{2\pi i} \sum_{\{\mathbf{k}_l \}} \tilde{F}^{\alpha}_{xy, l} \, \hat{p}^{\alpha}_l(E_F)
\end{equation}
with an error $\pm \Delta \tilde{\mathfrak{C}}_\alpha(E_F)$ of
\begin{equation}
\label{eq:Delta_C}
\Delta \tilde{\mathfrak{C}}_\alpha(E_F)  =  \sqrt{\sum_{\{\mathbf{k}_l\}}(\Delta \hat{p}^\alpha_l(E_F) \tilde{F}^\alpha_{xy,l})^2}
\end{equation}
with confidence of at least $1-\epsilon$. We remark that controllable error bars are particularly important and valuable outside of the insulating regime, i.e.~where the Fermi energy cuts a partially filled energy band, as in this case the Berry conductivity is not quantized and can assume continuous non-integer values.

\section{\label{sec:application}Practical Application of the Algorithm}

In this section, we apply the algorithm to different models. We first start with the Haldane model, a two band model that can realize both topological and trivial phases. We then go to the Hofstadter model, a multi-band model characterized by non zero Chern number and finish with the BHZ model, a two band realistic model realizing a quantum spin Hall effect in condensed matter physics.

\subsection{\label{sec:models:haldane}The Haldane model}


The model proposed by Haldane in\cite{haldane-prl-61-2015} is a tight-binding Hamiltonian of spinless fermions on a honeycomb lattice, with dynamics governed by nearest-neighbor (N.N.) real-valued hopping term of amplitude $J$ and an imaginary next-to-nearest neighbor (N.N.N.) hopping term $J_2$ (see Fig.~\ref{fig:fig2}a). In addition, the fermions are exposed to an onsite staggering potential $\beta$, which induces a chemical potential difference between nearest-neighbors sites of the bi-partite hexagonal lattice ($\phi$ and $\psi$ sites). The model is exactly solvable and represents a paradigmatic model in the field of topological phases of matter, as it hosts a quantum AHE phase even in the absence of an external magnetic field. Recently, it has been proposed that the physics of this model could be observed experimentally in a quantum simulation with cold atoms in optical lattices \cite{alba-prl-107-235301}. 

The Hamiltonian of the system is given by
\begin{equation}
\label{eq:hladanemod}
H=- J \sum_{\langle i,j \rangle} c^\dagger_i c_j +i J_2 \sum_{\langle\langle i,j \rangle\rangle} \nu_{ij} c^\dagger_i c_j+ \beta \sum_i s_i c^\dagger_i c_i\text{,}
\end{equation}  
Here, $c^\dagger_i$ and $c_i$ are fermionic creation and destruction operators, $\nu_{ij}=\text{sgn}[(\mathbf{d}_1\times \mathbf{d_2})_z]$ and $s_{\phi,\psi}=\pm1$. The vectors $\mathbf{d}_1$ and $\mathbf{d}_2$ are oriented along the bonds of the hexagonal unit cell, as shown in Fig.~\ref{fig:fig2}a. The model can be readily solved by rewriting the Hamiltonian in terms of two-site basis cells $(\phi,\psi)$ (see e.g.~\cite{dauphin-pra-86-053618}) such that the hexagonal lattice becomes a triangular lattice of $(\phi,\psi)$ cells. In the Fourier space the Hamiltonian is then given by \cite{haldane-prl-61-2015}
\begin{equation}
H=\sum_{\mathbf{k}\in B.Z.}\hat{\Psi}^\dagger(\mathbf{k})\left(\begin{array}{cc}\beta-2J_2 f(\mathbf{k}) & -A^*(\mathbf{k}) \\A(\mathbf{k}) & -\beta+2J_2 f(\mathbf{k})\end{array}\right)\hat{\Psi}(\mathbf{k})\text{.}
\end{equation}
Here, $\hat{\Psi}^\dagger(\mathbf{k})=(c^\dagger_\phi(\mathbf{k}),c^\dagger_\psi(\mathbf{k}))$, $A(\mathbf{k})=\exp(i\mathbf{k}\cdot \delta_1)+\exp(i\mathbf{k}\cdot \delta_2)+\exp(i\mathbf{k}\cdot \delta_3)$ is expressed in terms of the vectors between nearest neighbors $\delta_1$, $\delta_2$ and $\delta_3$ and $f(\mathbf{k})=\sin[\mathbf{a}_1\cdot \mathbf{k}]+\sin[\mathbf{a}_3\cdot \mathbf{k}]+\sin[(\mathbf{a}_1+\mathbf{a}_2)\cdot \mathbf{k}]$ is expressed in terms of the lattice vectors $\mathbf{a}_{1}$ and $\mathbf{a}_{2}$ as shown in Fig.~\ref{fig:fig2} and defined in Appendix \ref{app:honeycomb}.

Diagonalization of the Hamiltonian readily yields the two-band energy spectrum
\begin{equation}
E_\pm(\mathbf{k})=\pm\sqrt{|A(\mathbf{k})|^2+(\beta-2J_2f(\mathbf{k}))^2},
\end{equation}
which is shown in Fig.~\ref{fig:fig2}b. 

\begin{figure}[t]
\begin{center}
	\includegraphics[scale=1]{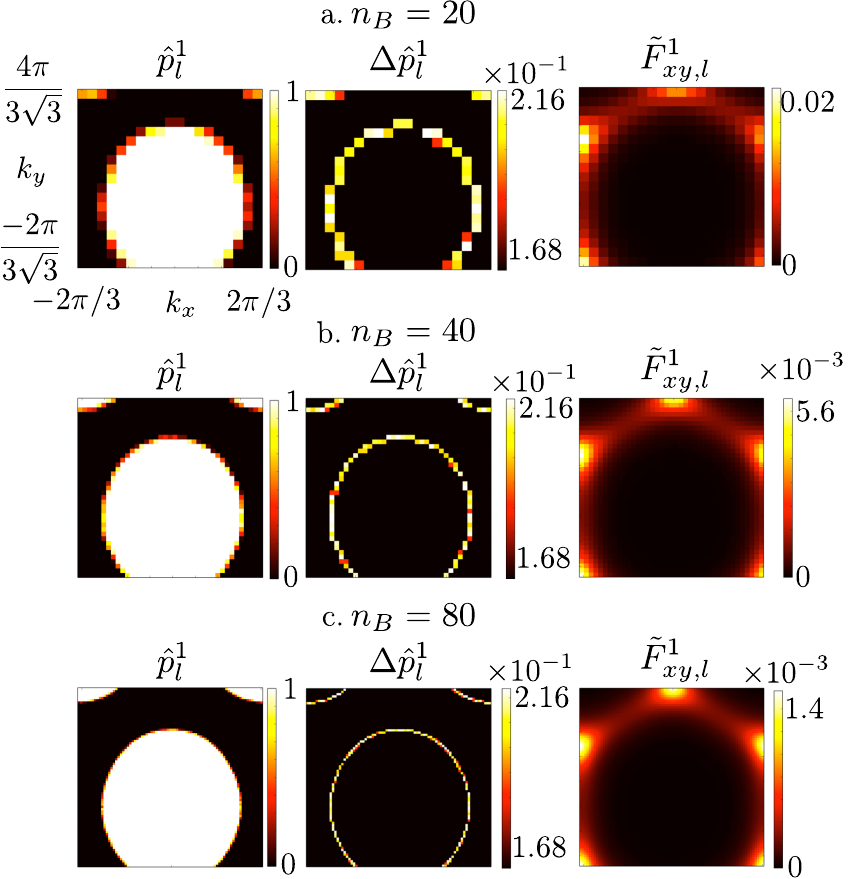}
	\caption{Central ingredients for the numerical calculation of the Berry conductivity : weight estimators $\hat{p}_l(E_F)$ (left column) with statistical errors (central column), and Berry curvature contributions $\tilde{F}_{xy}$ (right column). The rows show the numerical results for increasingly finer grids of the Brillouin zone: $n_B=20$ (upper), $n_B=40$ (central) and $n_B=80$ (lower row). The results are obtained for the Haldane model for the Fermi energy lying in the lower band at $E_F=-1.5J$, and for Hamiltonian parameters $J_2=0.1J$ and $\beta=0$, and a sampling of $n_R = 20$ random points per momentum space plaquette.}
\label{fig:fig3}
\end{center}
\end{figure} 

For $\beta = J_2 = 0$, the Hamiltonian corresponds to pure nearest-neighbor hopping of fermions with the characteristic spectrum exhibiting the two inequivalent Dirac cones \cite{bib:bena,castro-rmp-81-109}. A non-zero staggering potential $\beta \neq 0$ induces an imbalance of the fermion density on $\phi$ and $\psi$ lattice sites. The formation of a charge-density-wave phase is associated to the opening of a topologically trivial insulating gap in the spectrum. On the other hand, a strong enough N.N.N. hopping term $J_2$ opens a topologically non-trivial energy gap that signals the transition of the system into a AHE phase characterized by a non-zero Chern number. The size of the energy gap is determined by the formula $\Delta=2|\beta-3\sqrt{3}J_2|$,  and for $|\beta|<3\sqrt{3}|J_2|$ the system is in the topological phase. 

We will now illustrate the working principle of our algorithm by applying it step by step -- as schematically summarized in Fig.~\ref{fig:fig1}c -- to the Haldane model. To this end, we start by fixing the Hamiltonian parameters to $J_2=0.1J$, $\beta=0$, i.e.~deep in the topologically non-trivial phase. Next, we discretize the Brillouin zone (step 1), where we use for numerical convenience a rectangular-shaped B. Z. parametrization which is equivalent to the standard hexagonal form (see Appendix \ref{app:honeycomb} for details). 

\begin{figure*}[t]
\begin{center}
	\includegraphics[scale=1]{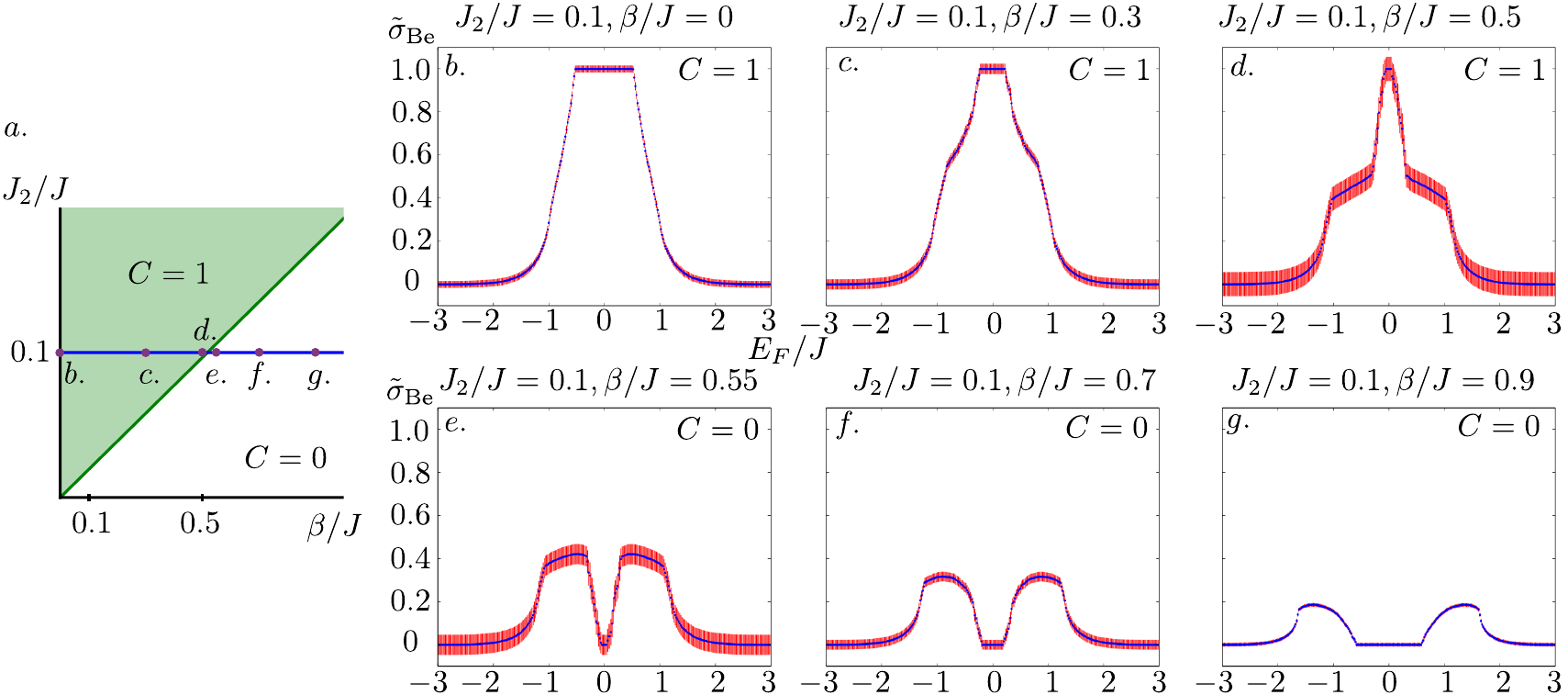}
	\caption{Numerically obtained Berry conductivity $\tilde{\sigma}_\text{Be}(E_F)$ in the Haldane model as the system undergoes the transition from the topologically nontrivial AHE phase (Chern number $C = 1$ for Fermi energies lying in the gap) to the trivial band insulator induced by the staggering potential (characterized by a Chern number $C = 0$ for Fermi energies in the gap). The closure and reopening of the gap as the transition from the topological to the trivial phase takes place is clearly reflected by the width of the conductance plateau around $E_F=0$, following the analytical $\Delta=2|\beta-3\sqrt{3}J_2|$ dependance. The results are obtained for a Brillouin zone grid parameter $n_B = 20$ and $n_R = 20$ random points per momentum space plaquette, and statistical error bars correspond to a confidence of 95 \% ($\epsilon = 0.05$).}
\label{fig:fig4}
\end{center}
\end{figure*}

Then, we compute the field strength $\tilde{F}_{xy}$ for each plaquette (step 2); the result is shown in the right column of Fig.~\ref{fig:fig3}. We fix the number of random points (we choose $n_R=20$) (step 3) and compute for each plaquette for $n_R$ randomly distributed momentum vectors $E_\alpha(\mathbf{k}_R)$ (step 4). Once the Fermi energy is fixed (step 5), here to a value of $E_F=-1.5J$ so that the Fermi energy level cuts the lower band, we compute the estimators for the weights $\hat{p}^\alpha_l (E_F)$ according to Eq.~(\ref{eq:probareaest}) (step 6). The values of the estimators $\hat{p}^\alpha_l (E_F)$ are shown in the left column of Fig.~\ref{fig:fig3}. The central column of the figure displays the associated statistical uncertainties $\Delta \hat{p}^\alpha_l(E_F)$, as determined in step 7 with the Wilson interval with modified boundaries and symmetrized (see Appendix \ref{app:detailsalgo}). As expected and desired, the statistical errors associated to plaquettes which correspond to regions that clearly lie above or below the Fermi energy are minimal. In constrast, the plaquettes at energies around $E_F$ which are cut by the Fermi surface, have higher values. Note that even for a very limited Monte Carlo statistics involving only $n_B = 20$ random points per momentum space plaquette, these uncertainty values are still much smaller than the upper bound of one. In fact, higher uncertainties and error bars for plaquettes around the Fermi surface reflect the physical fact that these are the plaquettes correspond to the regions in momentum space where small changes in the Fermi energy level can lead smaller or larger contributions of Berry curvature and thus to changes in the Berry conductivity. The central and lower row of Fig.~\ref{fig:fig3} show the weight estimators, uncertainties and Berry curvature contributions for larger values of $n_B$, illustrating how an increasingly finer grid of the Brillouin zone leads to an increased resolution and numerical precision.

Finally,  the estimated weights $\hat{p}^\alpha_l (E_F)$ and the Berry curvature contributions $\tilde{F}^\alpha_{xy, l}$ are combined to calculate the Berry conductivity (step 8) according to Eqs.~(\ref{eq:Berry_Conductivity_estim})  and (\ref{eq:partialchern}) with an associated error bar (step 9) as given by Eq.~(\ref{eq:Delta_C}). By applying the algorithm again for varying values of the Fermi energy, the Berry conductivity can be obtained as a function of the Fermi level energy. The obtained Berry conductivity is shown in Fig.~\ref{fig:fig4}a: starting from low conductivity values at the bottom of the lower energy band, the conductivity increases up to its plateau value of one for Fermi energies lying in the topological insulating gap, before it subsequently starts to fall off again once the Fermi energy reaches the upper band. 

To test the behavior of the algorithm when the system undergoes a phase transition from the topological AHE phase to the trivial insulating phase, we increase the Hamiltonian parameter $\beta$ to observe the competition of the N.N.N. hopping term with the staggering potential. The subplots in Fig.~\ref{fig:fig4} show the transition from the topologically non-trivial phase characterized by a Chern number of one to the topologically trivial charge-density-wave phase with a vanishing Chern number. The algorithm correctly captures the closing of the gap as well as the jump of the conductivity plateau-value as the phase transition takes place. We emphasize that the algorithm automatically takes into account the fact that at the phase transition the Berry curvature is highly localized at the Dirac points and thus concentrated in only few plaquettes - a fact that the algorithm signals in the form of larger error bars of the Berry conductivity in the parameter regime where the transition occurs. \\

\begin{figure}[t]
\begin{center}
	\includegraphics[scale=1]{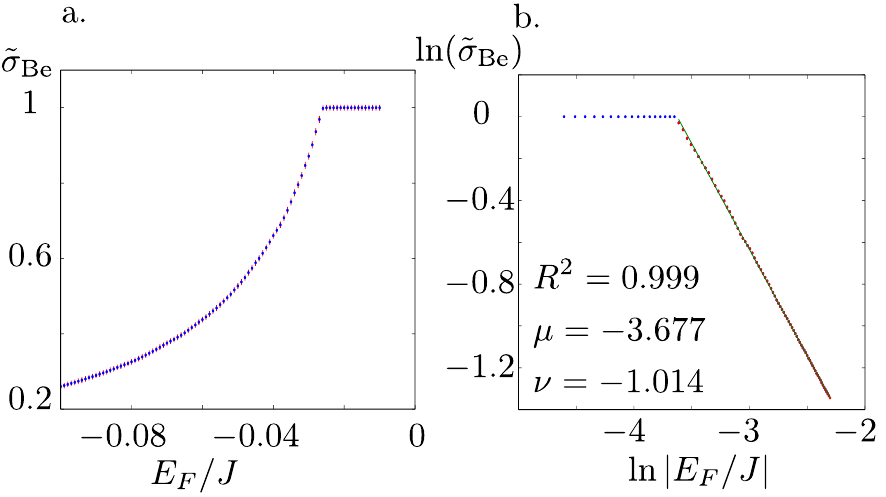}
	\caption{Berry conductivity in the Haldane model at Fermi energies in the vicinity of the gap (Hamiltonian parameters are fixed at $J_2=0.005 J$ and $\beta=0$ and for the parameters $n_B=450$ and $n_R=40$). \textbf{a.} Rapid increase of the conductivity to the plateau value, as the Fermi energy approaches the band gap. \textbf{b.} Double-logarithmic plot of the conductivity, $\ln(\sigma_{Be}(E_F)\,(h/e^2))=\nu \ln|E_F / J|+\mu$. A linear regression analysis of the numerical data yields the scaling exponent $\nu = - 1.014$ and $\mu = -3.677$ for a squared correlation coefficient $R^2=0.999$. These values coincide with the theoretically predicted values of $\nu = -1$ and $ \mu = (\ln 3 \sqrt{3} J_2 / J) =  -3.650$ around 1\%.}
\label{fig:fig5}
\end{center}
\end{figure} 

Finally, we apply the algorithm to the case where the system resides in the topological phase with a small topological gap opened. Here, the algorithm allows one to clearly verify numerically the $1/E_F$ power law dependence of the Berry conductivity for Fermi energies close to the gap. The $\sigma_{Be} (E_F) = (e^2/h) \, 3 \sqrt{3} J_2 / |E_F|$  behavior is predicted by the linear approximation of the spectrum around the Dirac points\cite{kane-prl-95-226801,PhysRevLett.97.106804,xiao-rmp-82-1959,Dyrda:2009p1760}. The results are shown and discussed in Fig.~\ref{fig:fig5}.

\subsection{\label{sec:models:hofstadter}The Hofstadter model}

\begin{figure}[t] 
\begin{center}
	\includegraphics[scale=1]{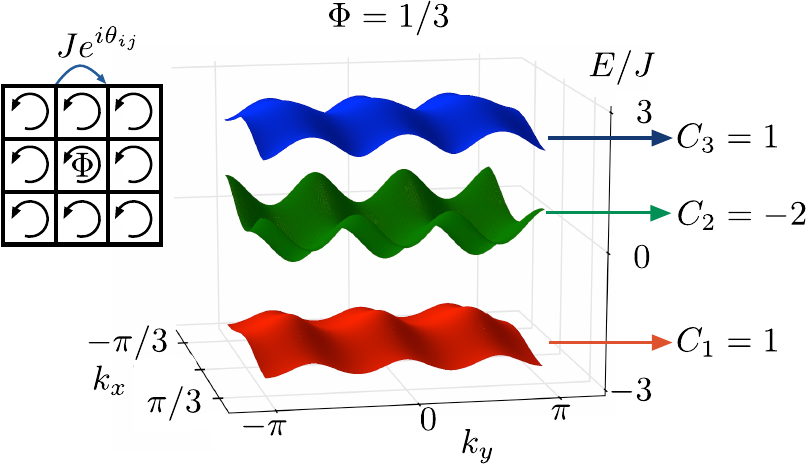}
	\caption{The Hofstadter model \cite{hofstadter-prb-14-2239} describes non-interacting spinless fermions on a square lattice under a magnetic flux $\Phi$ quanta per unit cell. For $\Phi=p/q$ a rational number, the energy spectrum of the bulk splits into $q$ sub-bands, as shown here for the case $\Phi=1/3$. Each band is characterized by a non-vanishing Chern number $C$. }
\label{fig:fig6}
\end{center}
\end{figure} 

Let us now apply the numerical algorithm to the Hofstadter model \cite{hofstadter-prb-14-2239}, which describes spinless fermions on a square lattice, subjected to a uniform magnetic field of magnetic flux quanta per unit cell $\Phi$. Only very recently, several groups have achieved to observe the characteristic physics including the fractal spectrum known as Hofstadter's butterfly in graphene superlattice systems \cite{ponomarenko-nature-497-594, dean-nature-497-598, hunt-science-340-1427}. This is complementary to ongoing experimental efforts to realize theoretical ideas \cite{jaksch-njp-5-56} on how to implement the fermionic Hofstadter Hamiltonian with cold atoms in optical lattices \cite{aidelsburger-prl-107-255301, PhysRevLett.111.185301,miyake-prl-111-185302, tarruell-nature-483-302}.

The Hamiltonian in second-quantized form is given by 
\begin{equation}
H=\sum_{\langle i,j \rangle}e^{i\theta_{ij}}c^\dagger_i c_j\text{,}
\end{equation}
where the sum is over nearest neighbor sites (see Fig.~\ref{fig:fig6}) and the phase factor $\exp(i\theta_{ij})$ corresponds to the Peierls substitution expressed in terms of the line integral over the vector potential along the link between two neighboring sites $i$ and $j$ of the square lattice. If $\Phi=p/q$ is a rational number, the energy spectrum of the bulk, described in the Fourier space, splits into q sub-bands, each one of them associated with a non-trivial integer-valued Chern number. 

\begin{figure}[t]
\begin{center}
	\includegraphics[scale=1]{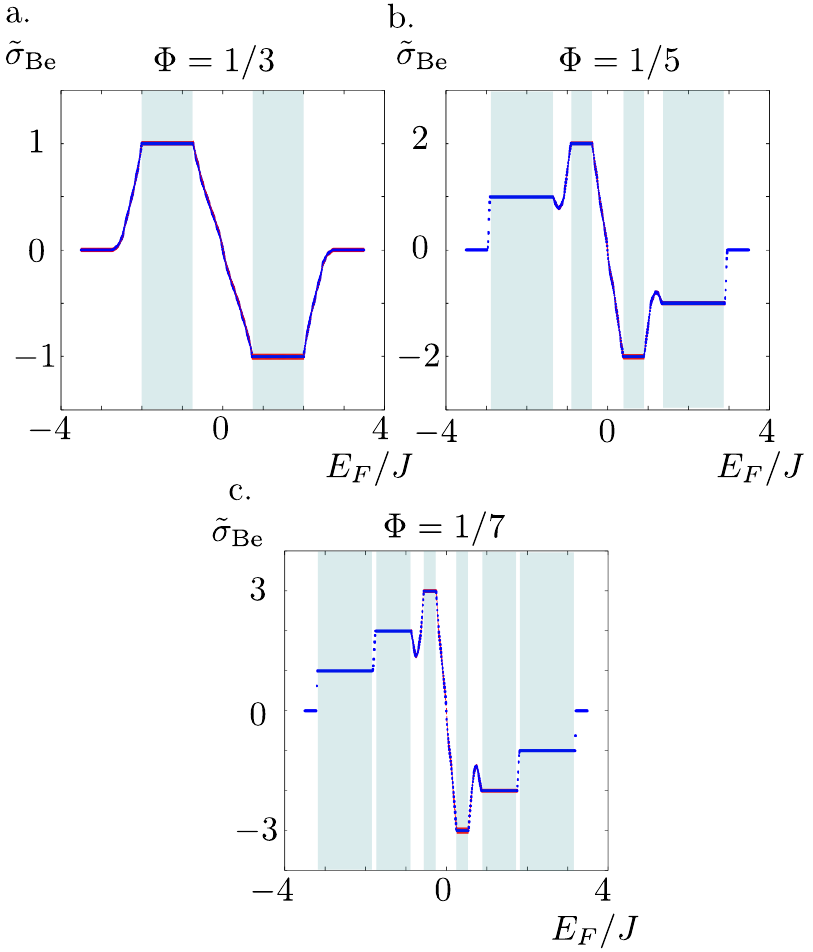}
	\caption{Numerical results for the Berry conductivity $\tilde{\sigma}_\text{Be}$ (blue points) as a function of the Fermi energy  for three values of the magnetic flux ($\Phi=$ 1/3, 1/5 and 1/7). The Brillouin zone has been discretized by a grid of $n_B=20$ with $n_R=20$ random points per momentum space plaquette. Statistical errors (red bars) correspond to a confidence of 95\% ($\epsilon = 0.05$).}
\label{fig:fig7}
\end{center}
\end{figure} 

Due to its multi-band structure the Hofstadter Hamiltonian can in general not be diagonalized analytically and thus represents an interesting testbed for the numerical algorithm. Fig.~\ref{fig:fig7} shows the numerical results for the Berry conductivity for different values of the flux per plaquette ($\Phi =$ 1/3, 1/5 and 1/7). For Fermi energies lying in the energy gap between bulk bands, the algorithm correctly reproduces the constant Berry conductivity, which corresponds to the sum of the Chern numbers of completely filled bands. Once the Fermi energy falls into a bulk band the Berry conductivity is no longer quantized. Whereas for $\Phi = 1/3$ the Berry conductivity interpolates monotonically between the gap plateau values, for $\Phi = 1/5$ the conductivity displays an interesting feature for Fermi energy values in the second band: instead of showing a monotonic growth, it first decreases to a minimum value, before starting to increase until it reaches the plateau dictated by the quantized value of the conductivity in the gap. The same phenomenon occurs, even more pronounced, in the third band of the spectrum for $\Phi = 1/7$. The small controlled statistical error bars of the numerical method ensure that the non-monotonic signature in the Berry conductivity is indeed a physical feature rather than a numerical artifact. 

\subsection{\label{sec:models:bhz}The BHZ model}

In 2005, it was suggested that the quantum spin Hall effect (QSHE) could possibly be observed in graphene \cite{kane-prl-95-146802,kane-prl-95-226801}, which however turned out to be impeded by too weak spin-orbit coupling in this system. Shortly later, a realization of the QSHE in HgTe/CdTe nanowell structures was proposed \cite{bernevig-science-314-1757} and experimentally realized only one year later \cite{koenig-science-318-766}: by varying the thickness of the different layers of the heterostructure, the material can exhibit a trivial insulating phase as well as a topological insulating phase, characterized by a $\mathbb{Z}_2$ topological invariant. The physics can be described by an effective Hamiltonian valid close to the $\Gamma$ point, derived by Bernevig, Hughes and Zhang (BHZ model) \cite{bernevig-science-314-1757,Konig:2008p1451}. The Hamiltonian is given by $4\times 4$ matrix in momentum space,

\begin{equation}
\begin{split}
H=\left(\begin{array}{cc}h(\mathbf{k}) & 0 \\0 & h^*(-\mathbf{k})\end{array}\right)\text{,}\\
h(\mathbf{k})=\epsilon(\mathbf{k})\mathbb{1}+d_i\sigma^i\text{,}
\end{split}
\end{equation}
where $\mathbb{1}$ is the two-dimensional identity matrix, $\sigma^i$ denote the Pauli matrices and 
\begin{equation}
\label{eq:parameter_list}
\begin{split}
\epsilon(\mathbf{k})=C-D(k_x^2+k_y^2)\text{,}\\
\mathbf{d}(\mathbf{k})=(Ak_x,-Ak_y,M(\mathbf{k})),\\
M(\mathbf{k})=M-B(k_x^2+k_y^2)\text{.}
\end{split}
\end{equation}
The parameters $A$, $B$, $C$, $D$ and $M$ depend on material properties as well as the thickness of the layers and can be computed numerically \cite{bernevig-science-314-1757,Konig:2008p1451}. 

\begin{figure}[t]
\begin{center}
	\includegraphics[scale=1]{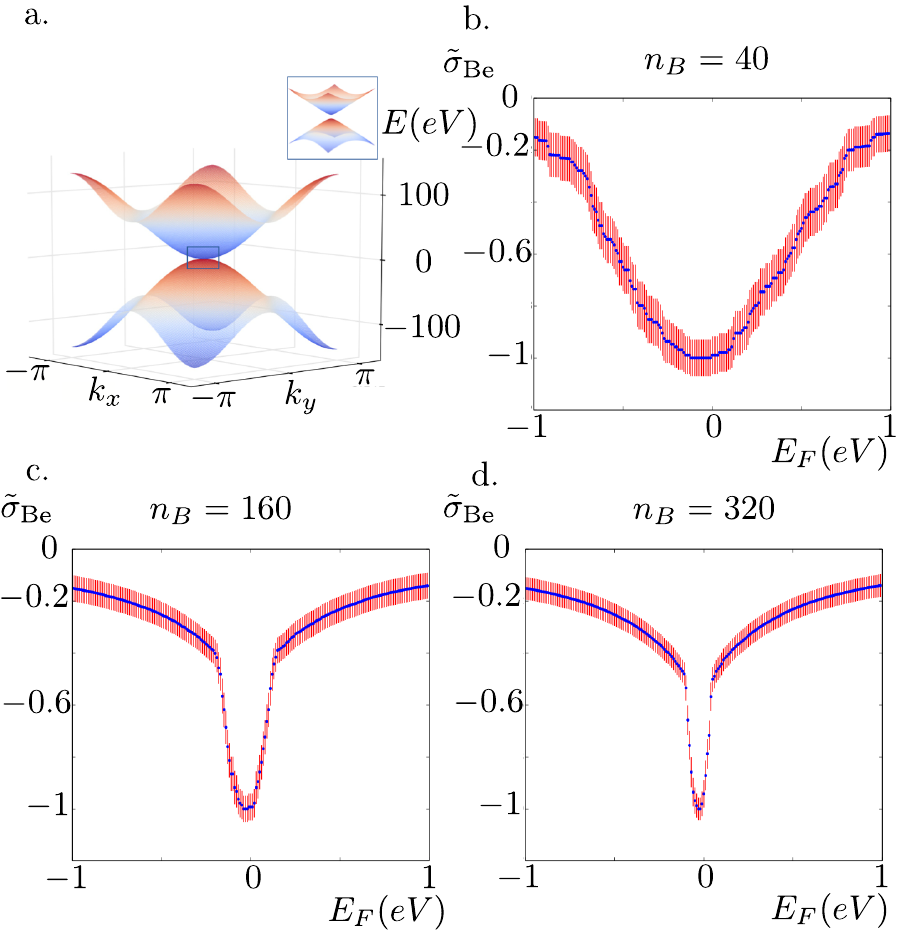}
	\caption{ Application of the algorithm to the BHZ model\cite{bernevig-science-314-1757,Konig:2008p1451}. a. Energy spectrum of the BHZ model exhibiting a small energy gap. The parameters of the model entering Eq.~(\ref{eq:parameter_list}) are chosen as $A=-3.42eV$, $B=-16.9eV$, $c=-0.0263eV$, $d=0.514eV$ and $M=-0.00686 eV$, as calculated in Ref. \cite{bernevig-science-314-1757}. The plots b, c and d show the numerically obtained Berry conductivity $\tilde{\sigma}_\text{Be}(E_F)$ for increasingly larger values of the momentum space resolution (grid sizes $n_B=$ 40, 160, 320). Error bars were obtained for $n_R=20$ and correspond to a confidence value of 95 \%.}
\label{fig:fig8}
\end{center}
\end{figure}

The Hamiltonian decouples into $2 \times 2$ blocks, and the spin conductivity can be written as the difference of the conductivity for each spin orientation and it makes thus sense to study the conductivity of one of the orientations. Here, we apply our algorithm to the BHZ model with parameters as calculated in \cite{bernevig-science-314-1757}. Figure~\ref{fig:fig8}a shows the energy spectrum that exhibits a small gap of $0.01eV$, which renders the computation of the Berry conductivity in the non-insulating regime more demanding. Figure~\ref{fig:fig8}b -- d show the numerical results for the Berry conductivity for increasingly finer grids of the Brillouin zone. 

Whereas even for the roughest grid studied ($n_B=40$) the algorithm correctly captures the qualitative behavior and the conductivity minimum value value of -1 for the Fermi energy lying in the shallow energy gap. However, as signaled by considerably large error bars, only few plaquettes contribute large values of Berry curvature to the conductivity. Thus, finer grids (see Fig.~\ref{fig:fig8}c and d with $n_B=160$ and $n_B=320$) are required to quantitatively correctly describe the conductivity behavior in the vicinity of the gap. This effect illustrates the importance of a high enough resolution, both numerically and in an experiment. As the algorithm qualitatively captures the behavior even for rather coarse-grained grids, this can be helpful to predict observations in the case of restricted experimental resolution, e.g.~originating from the finite size of optical lattices for cold atoms, or finite temperature constraints in solid state experiments. In Appendix \ref{app:conv}, error bounds for the conductivity which take into account a finite grid resolution are discussed in detail.

Finally, we remark that the BHZ model is an effective model valid close to the $\Gamma$ point, and thus the results of our analysis are also only valid in the vicinity of the energy gap. It is possible and will be an interesting extension of the present work to apply the numerical method to a more realistic, refined model which incorporates more information about the band structure of the system. 

\section{\label{sec:conclusions_and_outlook} Conclusions and Outlook}
In this work we have proposed and constructed a numerical algorithm to calculate the Berry conductivity in topological band insulators. The algorithm works for both the insulating case where the Fermi energy lies in the gap between two bulk bands as well as the situation where it lies within a band. The algorithm is gauge-invariant by construction, efficient and outputs the Berry conductivity with known and controllable error bars. We have successfully applied the algorithm to several paradigmatic models of topological quantum matter, including Haldane's model on the honeycomb lattice \cite{haldane-prl-61-2015}, the multi-band Hofstadter model \cite{hofstadter-prb-14-2239} and the BHZ model \cite{bernevig-science-314-1757} that describes the 2D spin Hall effect observed in CdTe/HgTe/CdTe quantum well compounds.

In addition to its applicability to topological insulators, the numerical method to compute the Berry conductivity for arbitrary values of the Fermi energy level can be applied to several other important problems: It can be used to study new phases of matter such as topological Fermi liquids \cite{haldane-prl-93-206602,castro-prl-107-106402, ritz-nature-497-231} which arise in interacting systems of fermions that realize a TI phase or an AHE phase. Mean field methods applied to these systems predict the existence of such phases \cite{Raghu-prl-100-156401, Weeks-prb-81-085105, dauphin-pra-86-053618}. Here, the efficient and controllable numerical method
for computing the Berry conductivity provides the appropriate observable to map out the possible topological phases of those systems with the desired accuracy \cite{grushin-prb-87-085136, araujo-prb-87-085109, hohenadler-jphys-25-143201}.

Recent experiments in which insulating phases \cite{tarruell-nature-483-302, aidelsburger-prl-107-255301} have been quantum simulated with cold atoms in optical lattices, provide another natural scenario where our new algorithm can be applied. Complementary to condensed matter systems, these experimental setups offer the possibility to study the intrinsic Berry conductivity in AHE systems under particularly clean and controllable conditions. Here, our algorithm can provide a precise observable to reliably and quantitatively distinguish symmetry protected topological phases from trivial phases and can predict some interesting features within the energy band. In fact, there have been proposed several ways to measure characteristic signatures of topological quantum phases in systems of cold atoms \cite{Price:2012p1561,Dauphin:2013p1761,PhysRevLett.111.120402,PhysRevLett.110.165304,2013arXiv1306.1190H}. In particular, recently several ways to measure the Berry conductivity in cold atoms experiments using time of flight measurements have been proposed \cite{alba-prl-107-235301,goldman-njp-15-013025,hauke-prl-109-145301}. 

An experimentally useful extension of our work would be to generalize our numerical method to the case of three dimensional topological insulators under time-reversal symmetry protecting conditions. Finally, it is an interesting question is how to generalize the controlled numerical method to an open quantum system scenario, such that it can be applied to topological insulators and topologically ordered systems coupled to an environment \cite{diehl-nphys-7-971, bardyn-arXiv:1302.5135,Viyuela:2013p1765, viyuela-prb-86-155140, viyuela-prb-88-155141}.

\section{Acknowledgments}
A. D. thanks the F.R.S.-FNRS Belgium for financial support and N. Goldman and P. Gaspard for support and valuable discussions. We acknowledge support by the Spanish MICINN grant FIS2009-10061,  FIS2012-33152, the CAM research consortium QUITEMAD S2009-ESP-1594, the European Commission
PICC: FP7 2007-2013, Grant No.~249958, and the UCM-BS grant GICC-910758.


\appendix

\section{\label{app:fukui}The FHS algorithm and the lattice gauge theory formulation}

The continuous Brillouin Zone is discretized by a two-dimensional lattice grid of $n_B$ points in each direction. For simplicity, we focus here on a rectangular grid, but the formalism can be readily extended to any polygonal grid \cite{phillips-annphys-prd-161-399}. The plaquettes of the momentum space lattice are then given by
\begin{equation}
\mathbf{k}_l=\mathbf{k}_{min}+i\mathbf{s}_{k_x}+j\mathbf{s}_{k_y}\text{,}
\end{equation} 
with 
\begin{align}
0 &\leq i,j \leq n_B-1\text{,} \nonumber \\
\mathbf{s}_{k_x} &=\delta k_x\mathbf{u}_{k_x}\text{,} \nonumber \\
\mathbf{s}_{k_y} &=\delta_{k_y}\mathbf{u}_{k_y}\text{,} \nonumber \\
 \delta_{k_x} &=(k_{xmax}-k_{xmin})/n_B, \nonumber \\
\delta_{k_y} &=(k_{ymax}-k_{ymin})/n_B\text{.}
\end{align}

The lattice field strength $\tilde{F}^\alpha_{xy}(\mathbf{k}_l)$ of band $\alpha$ on the grid is then defined in terms of the link variable $U_\mu (\mathbf{k})$ as 
\begin{equation}
\tilde{F}_{xy}(\mathbf{k}_l):= \ln[U_x(\mathbf{k}_l)U_y(\mathbf{k}_l+\mathbf{1}_{k_x})/(U_x(\mathbf{k}_l+\mathbf{1}_{k_y})U_{k_y}(\mathbf{k}_l))]\text{,}
\end{equation}
where $U_\mu(\mathbf{k})=\langle u(\mathbf{k})|u(\mathbf{k}+\mathbf{1}_\mu) \rangle$. If the \emph{admissibility condition} $|\tilde{F}_{xy}(\mathbf{k}_l)| < \pi$ is satisfied \cite{bib:fukui, phillips-annphys-prd-161-399}, the lattice gauge theory corresponds to the continuous gauge theory \cite{bib:fukui, phillips-annphys-prd-161-399} and one can write:
\begin{equation}
|F_{xy}(\mathbf{k}_l)| \delta k_x\delta k_y	\simeq |\tilde{F}_{xy}(\mathbf{k}_l)|
\end{equation}
Based on these Berry curvature contributions, the Chern number can be computed as
\begin{equation}
\tilde{C}=\frac{1}{2\pi i}\sum_{\mathbf{k_l}}\tilde{F}_{xy}(\mathbf{k}_l)\text{.}
\end{equation}

\section{\label{app:detailsalgo}Choice and the computation of the statistical error}

In this section, we present the concept and the details of a confidence interval (C.I.) to characterize the statistical uncertainty of the estimated weights $\hat{p}^\alpha_l$, as defined in Eq.~(\ref{eq:probareaest}). For simplicity of the notation, we suppress the band index $\alpha$ and momentum index $l$ in the following. 

The problem of estimating the weights corresponds to determining the unknown, though fixed probability value $p$ of a binomial distribution $\mathcal{B}(n_R,\,p)$, based on the outcome of $n_R$ trials. The probability to observe $k$ of the $n_R$ enquiries the value +1 is given by
\begin{equation}
P(X = k)=\frac{n_R!}{k!(n_R-k)!}p^k(1-p)^k\text{.}
\end{equation}

The goal is to associate a C.I. of a width much smaller than one to the estimated value $\hat{p}$, such that the true value $p$ lies with a probability $1- \epsilon$ inside the C.I. There are several ways to define the C.I., and we will in the following outline the advantages and inconveniences of some of them to motivate the necessity to adopt a simple and appropriate one that we use in our algorithm. To characterize and compare the quality of different conventions for the C.I, it is convenient to introduce the \textit{coverage probability}: it corresponds to the \emph{effective probability} to be inside the C.I. and can be compared to the expected probability $1-\epsilon$. As a guiding principle, a "good" C.I. is an interval with $p_\text{cov}\simeq 1-\epsilon$. On the contrary, for $p_\text{cov}<1-\epsilon$, the C.I. is "bad" as the statistical "guaranteeing functionality" of the interval fails. The other case $p_\text{cov}>1-\epsilon$ is not dramatic in our context as this implies that the true value of the estimated quantity $p$ actually lies in the C.I.~with a probability even higher than the targeted value of $1-\epsilon$. 

The construction of C.I. is based on the central limit theorem, which can be used to prove the convergence of the Binomial distribution to a normal distribution $\mathcal{N}$, in our case:
\begin{equation}
\sqrt{n_R} \frac{\hat{p}_l-p}{\sqrt{p(1-p)}} \rightarrow \mathcal{N}(0,1) \text{ when } n_R\rightarrow \infty\text{.}
\end{equation}

The central limit theorem and the definition of the C.I. of a normal distribution with an expected probability $1-\epsilon$ permits us to write the C.I. of $\hat{p}_l$ as a self-consistent equation in terms of $p$:
\begin{equation}
\label{app:normalCI}
p=\hat{p}_l\pm z_{\epsilon/2}\sqrt{\frac{p(1-p)}{n_R}}\text{,}
\end{equation}
where $z_\epsilon$ is the quantile function of the normal distribution\cite{cowan1998statistical}. 

A first way to define a C.I. is by maximizing the second term of the sum, yielding
\begin{equation}
\label{eq:maxint}
p=\hat{p}_l\pm \frac{z_{\alpha/2}}{2\sqrt{n_R}}
\end{equation}
for $p=1/2$. This relation highlights the typical $1/\sqrt{n_R}$ dependence of the statistical error and can be used to provide a rough estimate of the size of the C.I. in terms of $n_R$. However, as the length of the interval does not longer depend on the estimated value $\hat{p}_l$ itself, it does not satisfy our requirement. It will have a coverage probability $p_\text{cov}>1-\epsilon$ and would output error bars that overestimate the actual uncertainty of the observable of interest.

\begin{figure}[h]
\begin{center}
	\includegraphics[scale=1]{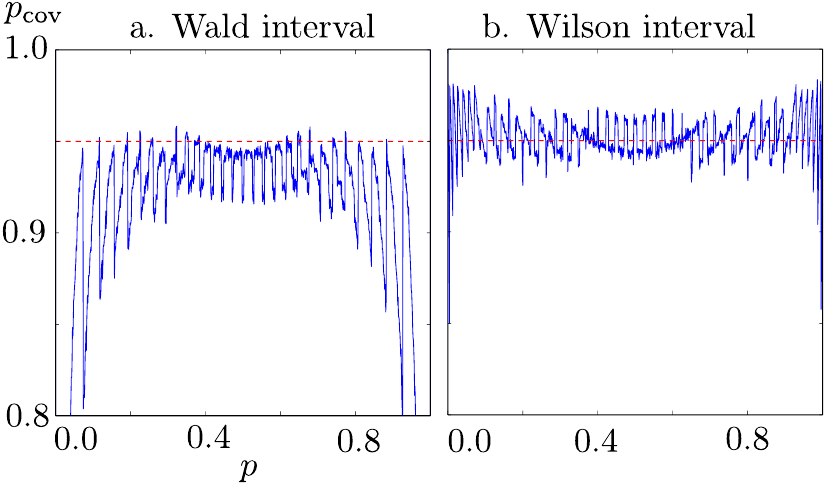}
	\caption{Plot of the coverage probability according to the Wald interval (a.) and the Wilson interval (b.) of a binomial distribution $\mathcal{B}(40,p)$ with an expected probability of $1-\epsilon=0.95$. The calculations have been done using $10000$ samples.}
\label{fig:app1}
\end{center}
\end{figure} 

Another commonly used C.I. ~is constructed using the approximation $p(1-p)\simeq \hat{p}_l(1-\hat{p}_l)$ in Eq.~(\ref{app:normalCI}), thereby replacing the unknown "true" value by the estimator value, so that 
\begin{equation}
\label{eq:waldint}
p=\hat{p}_l\pm z_{\epsilon /2} \sqrt{\frac{\hat{p}_l(1-\hat{p}_l)}{n_R}}\text{,}
\end{equation} 
This C.I. is known as the \emph{Wald interval} \cite{brown-stat-science-16-101}. Despite its simplicity, this convention suffers from several problems: for $\hat{p}_l \simeq 0$ or $\hat{p}_l \simeq 1$, the Wald interval shrinks to zero, implying a bad a coverage probability for $p$-values close to one or zero. As discussed by Brown \emph{et al.}\cite{brown-stat-science-16-101}, a series of criteria has been used in the literature to test the region of validity of this C.I. However, these criteria can be misleading and do not always characterize correctly the C.I. 

\begin{figure}[h]
\begin{center}
	\includegraphics[scale=1]{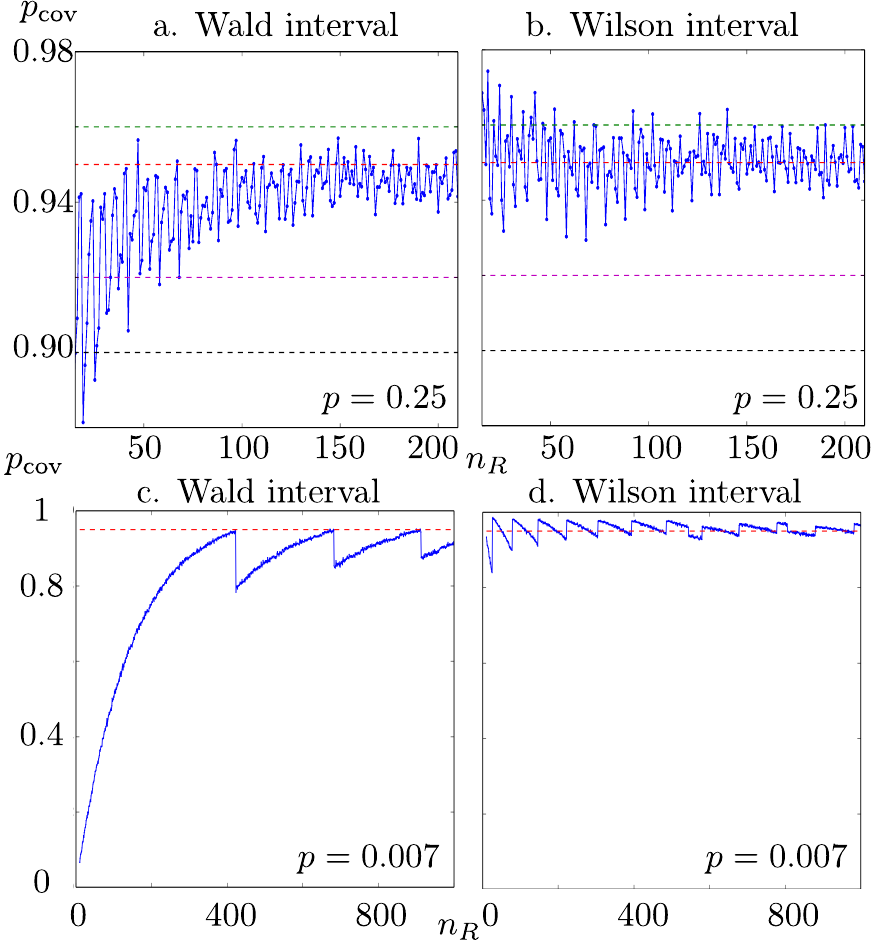}
	\caption{Plots a.~and b.~show the coverage probability according to the Wald and to the Wilson intervals of a binomial distribution $\mathcal{B}(n,0.25)$ with a probability $1-\epsilon=0.95$, where $20\leq n\leq200$. Plots c.~and d.~show the coverage probability according to the Wald and to the Wilson intervals of a binomial distribution $\mathcal{B}(n,0.007)$ with a probability $1-\epsilon=0.95$, where $10\leq n\leq1000$. The calculations have been done using $10000$ samples.}
\label{fig:app2}
\end{center}
\end{figure}

In Fig.~\ref{fig:app1}a we illustrate this problem for a fixed value of $n_R=40$ and by computing the coverage probability of the C.I. in terms of the value of $p$ on $10 000$ samples. One notices at first glance the tendency of the curve to lie below the expected value of $1-\epsilon$. Although the C.I. works rather well for values of $p$ close to $p=0.5$, it captures only poorly the situation at values close to the boundaries. Finally, the curve has a fast and significant oscillating behavior which gives rise to the phenomenon of so-called \emph{lucky/unlucky numbers}: when increasing slightly the probability $p$, the coverage probability jumps from a good $p_\text{cov}$ to a poor $p_\text{cov}$ value as it is the case for instance around $p=0.8$ in the shown example. The couple $(p,n_R)$ defines the lucky/unlucky numbers. In Fig.~\ref{fig:app2}a we fix the value $p=0.25$ and vary the value of $n_R$. Here one observes also significant fluctuations that are only stabilized at larger values of $n_R$. This effect becomes is even more striking at small $p$, as illustrated in the Fig.~\ref{fig:app2} c. where a fixed value of $p=0.007$ has been chosen: under an increase of $n_R$, the C.I. seems to converge to a favorable value of $p_\text{cov}$ until reaching $n_R=423$ where $p_\text{cov}$ suddenly drops from $0.94$ to $0.78$. We thus exclude the Wald interval as a candidate to construct the C.I. for the $\hat{p}^\alpha_l$ estimators in our algorithm. 

Most of the mentioned problems can be avoided if the approximation $p(1-p)\simeq \hat{p}_l(1-\hat{p}_l)$ is not applied in Eq.~(\ref{app:normalCI}). Instead, one can exactly solve Eq.~(\ref{app:normalCI}), which is a quadratic equation for $\hat{p}$. This yields the so-called \emph{Wilson interval} \cite{citeulike:3807920,brown-stat-science-16-101}
\begin{align}
p_{\text{max, min }} = \left( \hat{p}_l+\frac{z^2_{\epsilon/2}}{2n_R}\pm z_{\epsilon/2}\sqrt{\frac{\hat{p}_l(1-\hat{p}_l)}{n_R}+\frac{z^2_{\epsilon/2}}{4n_R^2}} \right) \Big/ \left( 1+\frac{z^2_{\epsilon/2}}{n_R} \right) \nonumber \\
\end{align}

As illustrated in the Fig.~\ref{fig:app1}b, the Wilson interval is much more stable and the coverage probability is oscillating around the value $1-\epsilon$. Figure~\ref{fig:app2} b. shows that the Wilson interval reaches rapidly and in a stable way the expected value $1-\epsilon$. The only problem still to be cured is at the boundaries, at $p$-values around zero or one, where the coverage probability drops. Figure~\ref{fig:app2} d. illustrates the convergence at small $p$, here fixed to $p=0.007$ and indicates that the effect of lucky/unlucky numbers is much less important than for the Wald interval. Brown \emph{et al.}\cite{brown-stat-science-16-101} propose to replace the lower (upper) boundary of the C.I. obtained by the normal approximation by a lower (upper) boundary obtained from a Poisson approximation for small (big) values of $\hat{p}$. This indeed stabilizes the behavior of the C.I. even close to the boundaries but complicates the expression of the C.I. Here, we propose a simpler patch, which has the same desired effect: we use the following replacement:
\begin{equation}
\begin{split}
&p_\text{min}=0 \text{ if } \hat{p}_l=x/n_R,x=0,1,2 \text{,}\\
&p_\text{max}=1 \text{ if } \hat{p}_l=x/n_R,x=n_R,n_R-1,n_R-2\text{,}
\end{split}
\end{equation}
including $x=3$ and $x=n_R-3$ if $n_R>40$. Finally, merely for convenience to obtain symmetric error bars, we symmetrize the C.I.~around $\hat{p}$ by choosing a width which corresponds to twice the value of $\text{max}\{ p_{\text{max}} - \hat{p}, \hat{p} - p_{\text{min}} \}$. While keeping the C.I.~narrow, this only leads to a modest over-estimation of the actual uncertainty of the estimator. 

The C.I.~interval defined in this form has a simple analytical form in combination with a good coverage probability, even for small $n_R$\cite{brown-stat-science-16-101}. We will use this construction of the C.I.~in the Monte Carlo sampling part of the algorithm, and refer to it as \textit{Wilson interval with modified boundaries} in the main text.

\section{\label{app:honeycomb} Properties of the honeycomb lattice}

\begin{figure}[h]
\begin{center}
	\includegraphics[scale=1]{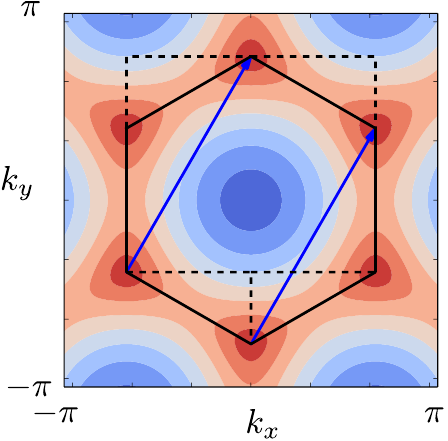}
	\caption{The hexagonal-shaped Brillouin zone is equivalent to a rectangle, obtained by a translation of two triangles (dashed line) in the direction of the basis vectors of the reciprocal lattice $\mathbf{b}_1$ and $\mathbf{b}_2$.}
\label{fig:app3}
\end{center}
\end{figure} 

The reciprocal vectors are $\mathbf{b}_1=(2\pi/3,2\pi/\sqrt{3})$ and $\mathbf{b}_2=(-2\pi/3,2\pi/\sqrt{3})$. The equivalence between the hexagonal Brillouin zone and the rectangular area used in the computation is shown in Fig.~\ref{fig:app3}: the lower left triangle can be translated along $\mathbf{b}_1$ and the lower right triangle can be translated along the $\mathbf{b}_2$.

\section{\label{app:seff} Effect of the choice of the resolution and the choice of the number of random points}

\begin{figure}[h]
\begin{center}
	\includegraphics[scale=1]{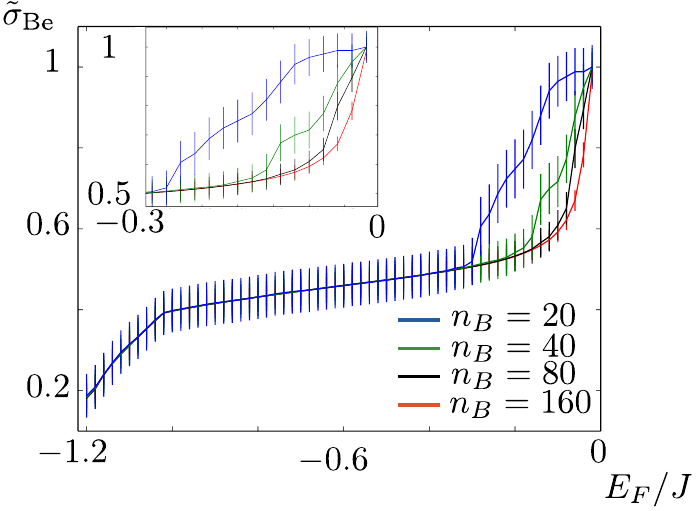}
	\caption{Plot of the numerically estimated Berry conductivity $\tilde{\sigma}_\text{Be}$ with error bars for four different values of the discretization number of the momentum space grid, $n_B=$ 20, 40, 80 and 160, for a system with parameters $J_2=0.1J$, $\beta=0.5J$ and $n_R=20$. The inset shows a zoom into the region $-0.3J\leq E_F \leq 0$. }
\label{fig:app4}
\end{center}
\end{figure} 

In this section, we illustrate the importance of an appropriate momentum space resolution, parametrized by the discretization number $n_B$. Figure~\ref{fig:app4} presents a zoom of Fig.~\ref{fig:fig4}c of the main text, showing the numerically estimated Berry curvature for different grids $n_B$. One finds that all graphs have the same behavior until reaching a value around $E_F=-0.33J$. There, the behavior of the estimator of the Berry curvature becomes jerky. This is a characteristics which shows up when some few momentum-space plaquettes have an important Berry curvature contribution. The error bars are signal this effect. The situation improves for increasing values of $n_B$: the curves converging to one sharp curve, showing that the main contribution of the Berry curvature stems from states with an energy close to zero, and the error bars decrease significantly.

\begin{figure}[h]
\begin{center}
	\includegraphics[scale=1]{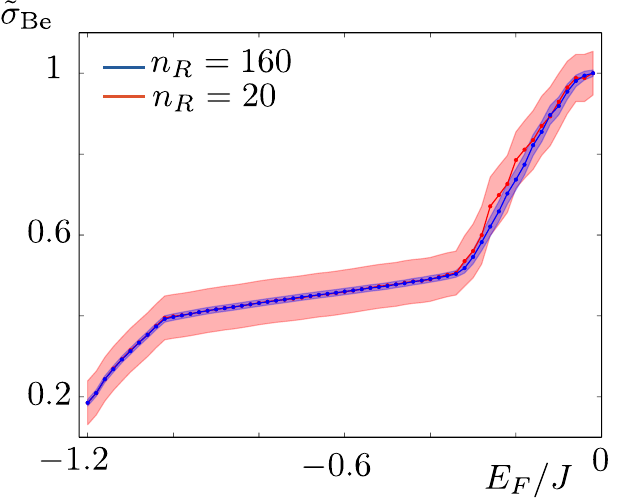}
	\caption{Numerically estimated Berry conductivity $\tilde{\sigma}_\text{Be}$ with error bars for two values of the number of random points $n_R=20$ and 160 for a system with $J_2=0.1J$, $\beta=0.5J$ and $n_B=20$. As expected, the computation with $n_R=160$ is much more precise, resulting in significantly smaller error bars. Note that as desired the $n_R=160$ curve is entirely comprised in the region spanned by the error bars of the computation with $n_R=20$.}
\label{fig:app5}
\end{center}
\end{figure} 

Another way to reduce the size of the error bars is to increase the value of $n_R$, the number of random points used to compute $\hat{p}_l$ in each plaquette. Figure~\ref{fig:app5} displays the estimated Berry conductivity for a fixed value $n_B=20$ and for the two values $n_R=20$ and $n_R=160$. As expected, the curve for $n_R=160$ is much more stable than the curve obtained for $n_R=20$: we see here a better interpolation in terms of the Fermi energy at this resolution. We emphasize the fact that the curve corresponding to $n_R=160$ is contained completely in the region spanned by the error bars of the $n_R  = 20$. This is an important point of the chosen construction of the confidence interval, as described in Appendix \ref{app:detailsalgo}. 

\section{\label{app:errbar} Importance of the choice of the error bars}

\begin{figure}[h]
\begin{center}
	\includegraphics[scale=1]{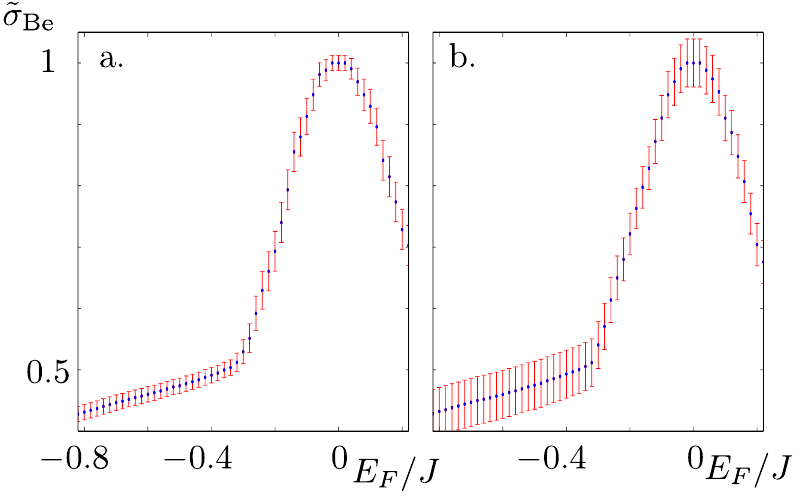}
	\caption{Plot of the conductivity for a system with $J_2=0.1J$, $\beta=0.5J$ and for $n_B=20$, $n_R=100$. The error bars are computed using the symmetrized Wilson interval with modified boundaries (left) and the C.I. defined in Eq.~(\ref{eq:maxint}). The sensitivity of the Wilson interval on the value of the Fermi energy $E_F$ is clearly visible.}
\label{fig:app6}
\end{center}
\end{figure}

In this section, we compare the Wilson interval with modified boundaries with the C.I. defined in Eq.~(\ref{eq:maxint}) by examining the final error interval obtained in the Haldane model using both methods. We work here with $J_2=0.1J$, $\beta=0.5J$ such that the Berry curvature is really sharp and localized. The Figure \ref{fig:app6} shows the results for both types of error interval with the parameters $n_B=20$, $n_R=100$. The error interval as obtained by using the Wilson interval (see Appendix \ref{app:detailsalgo}) captures correctly the fact that the main contribution to the Berry curvature is strongly localized in momentum space. This gives rise to an increased statistical error in the energy region in which the Fermi energy crosses plaquettes with a large contribution to the Berry curvature. The error obtained with the other C.I. , presented in Fig.~b., is constant and independent of the value of the Fermi energy and is thus not indicating the region where the Fermi energy crosses plaquettes with a large contribution to the Berry curvature. This point illustrates the choice of the Wilson interval to construct the main algorithm. 

\section{\label{app:conv} Error bound due to the grid resolution}

The algorithm introduced in this paper allows one to compute the Berry conductivity with controllable statistical error bars for a given grid resolution of the discretization of the Brillouin zone. 

The choice of this grid can be dictated either by physical constraints of the problem, such as e.g. by the finite size of the considered lattices in real space, for instance in experiments with cold atoms in optical lattices, or by limited numerical resources. In any case, it is important and highly desirable to be able to characterize the error due to the grid resolution. More precisely, for a given grid, the aim is to provide an upper bound on by how much  the Berry conductivity might at most change if the grid were chosen even finer. In this section, we address this problem and construct a tight error bound in terms of the grid resolution. In particular, this bound will require no inputs except the Berry curvature contributions of the small momentum space plaquettes, which in any case need to be determined (see step 2. in Fig. 1c) in the course of computing the Berry conductivity by our algorithm.

\begin{figure}[h]
\begin{center}
	\includegraphics[scale=1]{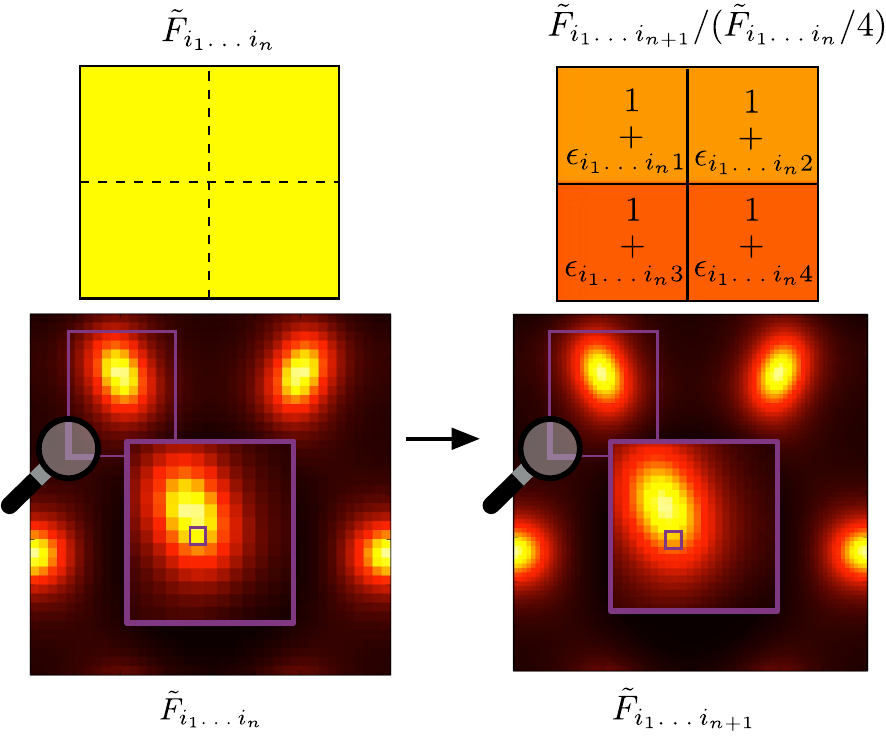}
	\caption{The Figure shows two grid levels $n$ and $n+1$ for the Haldane model with parameter $J_2/J=0.5,\beta/J=0$. The plaquette at the iteration $n$ is now described by four plaquettes at the iteration $n+1$. The $\epsilon_{i_1…i_{n+1}}$ are characterizing the difference with the homogeneous case. This is illustrated for one plaquette close to the Dirac cone.}	
\label{fig:app7}
\end{center}
\end{figure} 

We start with a discretization of the B.Z. into $L\times L$ plaquettes such that the sum $\sum_{i_1=1}^{L^2} \tilde{F}_{i_1}$ over the whole plaquettes is equal to the Chern number- in this section we are omitting the band index for notational simplicity-. Consider then finer and finer grids, where the $n$-th level grid contains $c^{n-1} L \times c^{n-1} L$ plaquettes : at each iteration, the previous plaquette is split into $c^2$ new small plaquettes as illustrated in Fig. \ref{fig:app7} which presents two successive grid levels with $c=2$.

The Berry conductivity at the iteration \textit{n} is written as:

\begin{equation}
\mathfrak{\tilde{C}}^{(n)}=\sum_{i_1=1}^{L^2}\sum_{i_2=1}^{c^2}…\sum_{i_n=1}^{c^2} \tilde{F}_{i_1…i_n}p_{i_1…i_n}\text{,}
\end{equation}

where 

\begin{equation}
\tilde{F}_{i_1…i_n}=\frac{1}{c^2}\tilde{F}_{i_1…i_{n-1}}(1+\epsilon_{i_1…i_n})
\end{equation}

and 

\begin{equation}
\frac{1}{c^2}\sum_{i_n=1}^{c^2}p_{i_1…i_n}=p_{i_1…i_{n-1}}\text{.}
\end{equation}

The parameters $\epsilon_{i_1…i_n}$  quantify the non-homogeneous contribution of the Berry curvature of the subplaquette in terms of the Berry curvature of the plaquette of the previous iteration.

The error bound at the iteration \textit{n},

\begin{equation}
\Delta \mathfrak{\tilde{C}}_\infty^{(n)}:= \sum_{m=0}^{\infty} \Delta \mathfrak{\tilde{C}}^{(n+m)}\text{,}
\end{equation}

is defined as the sum over all the relative errors $\Delta \mathfrak{\tilde{C}}^{(n+m)}=\mathfrak{\tilde{C}}^{(n+m)}-\mathfrak{\tilde{C}}^{(n+m+1)}$ between the iterations $n+m$ and $n+m+1$.

To compute the upper bound of this quantity, we should do an assumption about the smoothness of the Berry curvature; we expect that when the coarse graining is sufficiently fine after $n_0$ iterations, the Berry curvature becomes smoother at each iteration. Formally, we assume that there exists an $n_0$ and a parameter $0 < q < 1$ such that for $n > n_0$: 

\begin{equation}
\label{eq:app:conv:massump}
\epsilon^{(n)}:= \max_{i_1…i_n} |\epsilon_{i_1…i_n}|\leq q \max _{i_1…i_{n-1}}|\epsilon_{i_1…i_{n-1}}|\text{.}
\end{equation}

Note that it is essential to numerically verify for a given model and set of Hamiltonian parameters that this natural assumption is indeed fulfilled, and to determine from which $n_0$ on - see also examples below.

Using the last inequality, it is straightforward to derive an upper bound of the Berry conductivity $C^{(n+1)}$ and of the relative error $\Delta C^{(n)}$ in terms of $C^{(n)}$:

\begin{align}
&\mathfrak{\tilde{C}}^{(n+1)}\leq (1+q \epsilon^{(n)}) \mathfrak{\tilde{C}}^{(n)}\text{,}\\
&\Delta \mathfrak{\tilde{C}}^{(n)}\leq q \mathfrak{\tilde{C}}^{(n)}\text{.}
\end{align}

By iterating, one finds the bound of $\Delta \mathfrak{\tilde{C}}^{(n+m)}$ in terms of $\mathfrak{\tilde{C}}^{(n)}$:

 \begin{equation}
\Delta \mathfrak{\tilde{C}}^{(n+m)}\leq q^{m+1} \epsilon^{(n)} (1+q\epsilon^{(n)})^m \mathfrak{\tilde{C}}^{(n)}\text{.}
\end{equation}

The total error $\Delta \mathfrak{\tilde{C}}_\infty^{(n)}$ is bounded by a geometric series which, when $q(1+q\epsilon^{(n)})< 1$, converges and can be bounded by the value

\begin{equation}
\label{eq:app:conv:errorbound}
\Delta \mathfrak{\tilde{C}}_\infty^{(n)}\leq \frac{q}{1-q(1+q\epsilon^{(n)})}\epsilon^{(n)} \mathfrak{\tilde{C}}^{(n)}\text{.}
\end{equation}

\begin{figure}[h]
\begin{center}
	\includegraphics[scale=1]{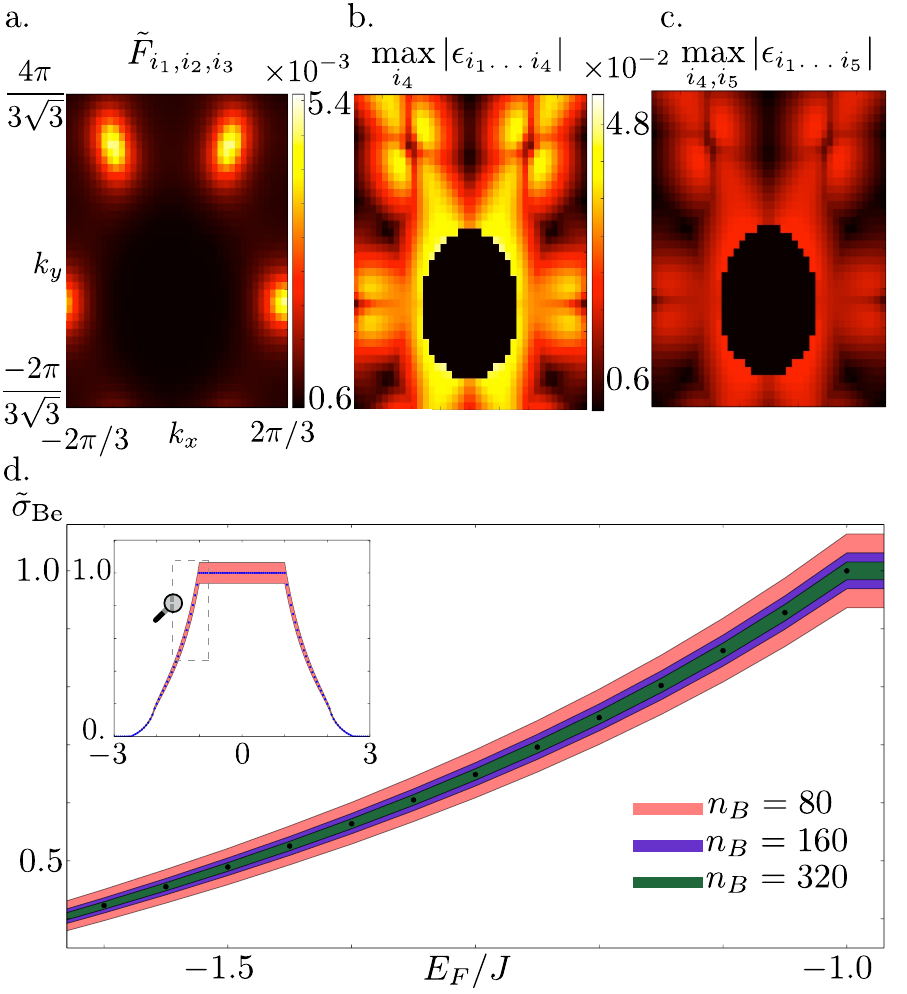}
	\caption{Determining the error bars due to the grid resolution. When the convergence regime is achieved, the Berry curvature $\tilde{F}_{i_1…i_n}$ becomes more homogeneous: each peak of the Berry curvature in Fig. a. is not anymore described by one plaquette. Furthermore, when we increase the grid size, the maximum of the deviation to the homogeneous case $\epsilon_{i_1…i_n}$ for each plaquette is getting smaller at each iteration as presented in the Figures b and c. Since the error due to grid comes from the  plaquettes with highest Berry curvature, we set a treshold of $10^{-2}\times |\max_{i_1…i_n}\tilde{F}_{i_1…i_n}|$ in Figure b and c and in the numerical computations. This information can be used to determine a maximal error interval for each grid resolution in the convergence regime. All these error bounds are containing the Berry conductivity computed for $n_B=320$ (black dots).}	
\label{fig:app8}
\end{center}
\end{figure} 

As an illustration we apply this formula to the Haldane model in the case of $J_2/J=0.5,\beta/J=0$. In this case, the Berry curvature is  smooth and the FHS algorithm already converges for $n_B=10$. We thus choose $L = n_B = 10$ for the first iteration. Figure \ref{fig:app8} a shows the Berry curvature after two iterations with $c=2$ (i.e. $n_B=40$). Following the exposed line of argument, we first verify numerically that we are in a convergence regime by testing the inequality \ref{eq:app:conv:massump} at each iteration. Figures \ref{fig:app8} b and c present the maximum of the  $\epsilon_{i_1…i_n}$ for two successive iterations in terms of the plaquette of the grid $n_B=40$. As expected, we find that the contribution of the $\epsilon$ at the next iteration level is getting smaller. In Table \ref{tab:table1}, we present more quantitative results for the different iterations for the $\epsilon_\text{max}^{(n)}$ at each iteration and the value of $q_n=\epsilon_\text{max}^{(n+1)}/\epsilon_\text{max}^{(n)}$. As it can be seen from the table, the value of $q_n=\epsilon^{(n+1)}_\text{max}/\epsilon^{(n)}_\text{max}$ is inferior to $1$ and by choosing $q=0.6$, we ensure that the relation \ref{eq:app:conv:massump} is satisfied. 

Next, the numerical factor of the error is obtained using the equation \ref{eq:app:conv:errorbound}. Figure \ref{fig:app8} displays the Berry conductivity for three successive iterations $n_B=80$, $160$, $320$ with error bounds. Here the parameter $n_R=200$ is chosen such that the statistical error is negligible. Otherwise, one should sum the statistical error to the error due to the grid.  The error bounds are becoming smaller at each iteration by a factor 2 and the conductivity converges rapidly with an error already of $0.03 \; \mathfrak{C}^{(n)}$ for $n_B=160$ and  $0.015\; \mathfrak{C}^{(n)}$ at $n_B=320$.

\begin{table}

    \begin{tabular}{| c | c | c || c | c | c | c | }
     \multicolumn{1}{c } { } \\
    \hline
     & \multicolumn{2}{ c|| }{$J_2/J=0.5$} & \multicolumn{2}{ c| }{$J_2/J=0.01$}\\
     \cline{2-3}\cline{4-5}
    $n_B=10\times 2^{n-1}$ & \hspace{2mm}$\epsilon^{(n)}_\text{max}$\hspace{2mm} & \hspace{2mm} $q_n$\hspace{2mm}  & \hspace{2mm} $\epsilon^{(n)}_\text{max}$\hspace{2mm}  & \hspace{2mm} $q_n$\hspace{2mm}  \\
    \hline
    $n=2$ & 0.197 & 0.419 & 0.693 & 0.847 \\
   
    $n=3$ & 0.087 & 0.471 & 0.588 & 0.713 \\
   
    $n=4$ & 0.041 & 0.496 & 0.419 & 0.516 \\
   
    $n=5$ & 0.020 & 0.496 & 0.216 & 0.472 \\
    
    $n=6$ & 0.010 & 0.497 & 0.102 & 0.489 \\
    
    $n=7$ & 0.005 & 0.499 & 0.049 & 0.491 \\
    \hline
   \end{tabular}
   
   \caption{The convergence of the Haldane model is studied for the parameter $\beta/J=0$ and two values of the parameter $J_2$: $J_2/J=0.5$ and $J_2/J=0.01$. When the regime of convergence is reached, the parameter $\epsilon^\text{max}_n$ is decreasingrapidly, and the ratio $q_n=\epsilon^{(n+1)}_\text{max}/\epsilon^{(n)}_\text{max}$ is converging to a constant value.}
   \label{tab:table1}
\end{table}

We have finally applied the algorithm on a more challenging case of the Haldane model with the parameters $J_2/J=0.01,\beta/J=0$. In this parameter regime, the band structure exhibits only a small energy gap between the two bands and the Berry curvature is very peaked at the Dirac cones. However, the FHS algorithm is already working at $n_B=10$. As it can be inferred from the right column of Table \ref{tab:table1}, the regime of convergence is only reached for a finer choice of the grid than in the case $J_2/J=0.50$ with at the beginning a slower decrease of $\epsilon_\text{max}^n$. However, at $n=4$ the system enters in a convergence regime and also in this case one can associate to the Berry conductivity upper error bounds due to the fine discretization of the Brillouin zone.


%

\end{document}